\documentclass[aps,pre,bibtex,twocolumn,notitlepage,superscriptaddress,10pt]{revtex4-2}

\usepackage{epsfig}
\usepackage{graphicx}
\usepackage{amsmath}
\usepackage{amssymb}
\usepackage{color}
\usepackage{hyperref}

\usepackage{ulem}

 \usepackage{xr-hyper}

\hypersetup{colorlinks=true,linkcolor=cyan,citecolor=blue}

\usepackage{lipsum}  
\usepackage{physics}
\newcommand{\ve}[1][K]{\mathbf{#1}}

\begin{document}

\title{Nanoscale Electroviscous Lift Force}


\author{Hao Zhang}
\thanks{These authors contributed equally to this work.}
\affiliation{State Key Laboratory of Structural Chemistry, Fujian Institute of Research on the Structure of Matter, Chinese Academy of Sciences, Fuzhou 350002, China}
\affiliation{Laboratoire Ondes et Mati\`ere d'Aquitaine, CNRS/University of Bordeaux, F-33400 Talence, France}

\author{Zaicheng Zhang}
\thanks{These authors contributed equally to this work.}
\affiliation{School of Physics, Beihang University, 100191 Beijing, China}

\author{Thomas Gu\'erin}
\email{thomas.guerin@u-bordeaux.fr}
\affiliation{Laboratoire Ondes et Mati\`ere d'Aquitaine, CNRS/University of Bordeaux, F-33400 Talence, France}

\author{Abdelhamid Maali}
\email{abdelhamid.maali@u-bordeaux.fr}
\affiliation{Laboratoire Ondes et Mati\`ere d'Aquitaine, CNRS/University of Bordeaux, F-33400 Talence, France}

\begin{abstract}
About forty years ago, it \textcolor{black}{was} predicted that a charged particle, moving parallel to a charged wall in an electrolyte,  should experience a lift force that, contrarily to electrostatic forces, is not screened at large distances. Up to now, such electroviscous lift force has not been directly measured. Here, we use Atomic Force Microscopy to directly measure the electroviscous lift force and quantify its dependency with the distance to the wall, the translation velocity or the particle’s size. Observing that existing theories exhibit large discrepancies with our experimental observations, we develop an analytical approach combining lubrication theory to a previously introduced formalism for small screening length. The experimentally observed lift forces are in good agreement with our theoretical predictions and reveal, for the first time, a saturation of the lift force for increasing velocities. Altogether, our results characterize, through direct measurements and analytical approach, the properties of electroviscous forces between charged particles in viscous electrolytes in non-equilibrium conditions. 
\end{abstract}

\maketitle

Electrolyte flows in confined geometries or near charged surfaces are fundamentally affected by the presence of electric double-layers (EDLs) that form at the interfaces \cite{lyklema2005fundamentals,robert2001fundamentals,stone2004engineering,bocquet2010nanofluidics}. 
Hydrodynamic flows disturb the ionic clouds next to the charged surfaces, resulting in the distortion of the EDL and inducing a dynamic coupling between viscous flow and charge distribution, giving rise to a range of electroviscous effects. These include ion transport in nanofluidic channels driven by surface charge \cite{stein2004surface}, reduction of diffusion of colloids near charged boundaries \cite{tawari2001electrical,eichmann2008electrostatically}, increased shear viscosity in charged colloidal suspensions \cite{horn2000hydrodynamic,rubio2004primary},  wall effect on charged particles during electrophoretic motion in microchannels \cite{KehJFM1985,KehJFM1988,KhairPRF2020,FernandezPRL2022}, and enhanced drag forces on charged particles moving near walls \cite{liu2018electroviscous,zhao2020electroviscous,rodriguez2022electroviscous,jin2022direct,cramail2025theory}.

About four decades ago, Prieve and co-workers \cite{alexander1987hydrodynamic,bike1990electrohydrodynamic,bike1992electrohydrodynamics,bike1995electrokinetic2} proposed an electrohydrodynamic mechanism for lubrication in electrolytes. Their theoretical calculations showed that for a charged sphere sliding near a charged surface, the flow-induced distortion of ion distributions generates Maxwell stresses, producing a repulsive electrokinetic lift force. While  electrostatic forces are   screened in electrolytes, this dynamic force, arising from electrohydrodynamic coupling, persists and decays algebraically with the distance to the surface. The calculated lift force increases with sliding velocity and fluid viscosity, but diminishes with increasing ionic strength. Notably, this force acts to prevent mechanical contact between surfaces \textcolor{black}{in sheared electrolytes}, suggesting a novel mode of electric lubrication in charged wet matter systems, \textcolor{black}{which could play a role in  the rheology of charged colloidal solutions or driven confined colloidal transport.}
It is worth noting that, for a neutral rigid sphere moving over a neutral rigid substrate at low Reynolds numbers, the reversibility of Stokes flow prohibits the emergence of a hydrodynamic force normal to the boundary. In the electrohydrodynamic mechanism \cite{alexander1987hydrodynamic,bike1990electrohydrodynamic,bike1992electrohydrodynamics,bike1995electrokinetic2} however, the normal stress includes a Maxwell stress contribution that generates a finite lift force. 
 This situation is analogous to the emergent lift forces in the elastohydrodynamics context that were recently observed for particles moving past soft interfaces \cite{saintyves2016self,rallabandi2018membrane,davies2018elastohydrodynamic,zhang2020direct,zhang2025direct,bureau2023lift}.
Subsequent theoretical work \cite{cox1997electroviscous,yariv2011streaming,schnitzer2012streaming} incorporated the role of hydrodynamic stresses and explicitly formulated the electroviscous formalism for gap distances larger than the Debye layer thickness. Explicit predictions were obtained for spheres or cylinders moving next to a plane wall \cite{warszynski1998electrokinetic,warszynski2000electroviscous,tabatabaei2006electroviscousSphere,tabatabaei2006electroviscousCylinder,schnitzer2016streaming}, in the limit of small velocities.   \textcolor{black}{ However, the different approaches lead to distinct predictions for the lift force, depending on the theoretical framework and regime considered.}

Despite this theoretical progress, experimental evidence for electroviscous lift remains limited. To date, all measurements of the electroviscous lift force have been \textcolor{black}{performed indirectly}, either by tracking particle heights under shear in the presence of gravity \cite{alexander1987hydrodynamic,bike1995electrokinetic,wu1996electrokinetic,Hollingsworth1996electrokinetic}  or by inferring forces through evanescent wave particle tracking in combined electroosmotic and pressure-driven flows \cite{cevheri2014lift}. \textcolor{black}{ The scarcity of  experimental data, and especially the absence of direct force measurements, has  prevented a quantitative assessment of the different theoretical predictions.} The colloidal-probe atomic force microscopy (AFM) technique has proven to be a powerful tool for measuring weak forces \cite{ducker1991direct,butt2005force}. Recently, this technique was successfully used to probe lift forces acting on spherical particles near soft interfaces \cite{zhang2020direct,zhang2025direct}.

In this work, we report direct and quantitative measurements of the electroviscous lift force acting on a spherical particle moving along a planar wall in an electrolyte solution. Using AFM, we measure the lift force as a function of separation distance, \textcolor{black}{ sliding velocity, and particle radius. Our measurements provide a direct test of existing theoretical predictions and reveal significant quantitative discrepancies.} We therefore develop an analytical approach based on Cox's formalism in the lubrication regime.  Our theory quantitatively matches  with observed lift forces and reveals a previously overlooked saturation of the force for high velocities.

\begin{figure}
\includegraphics[width=8cm]{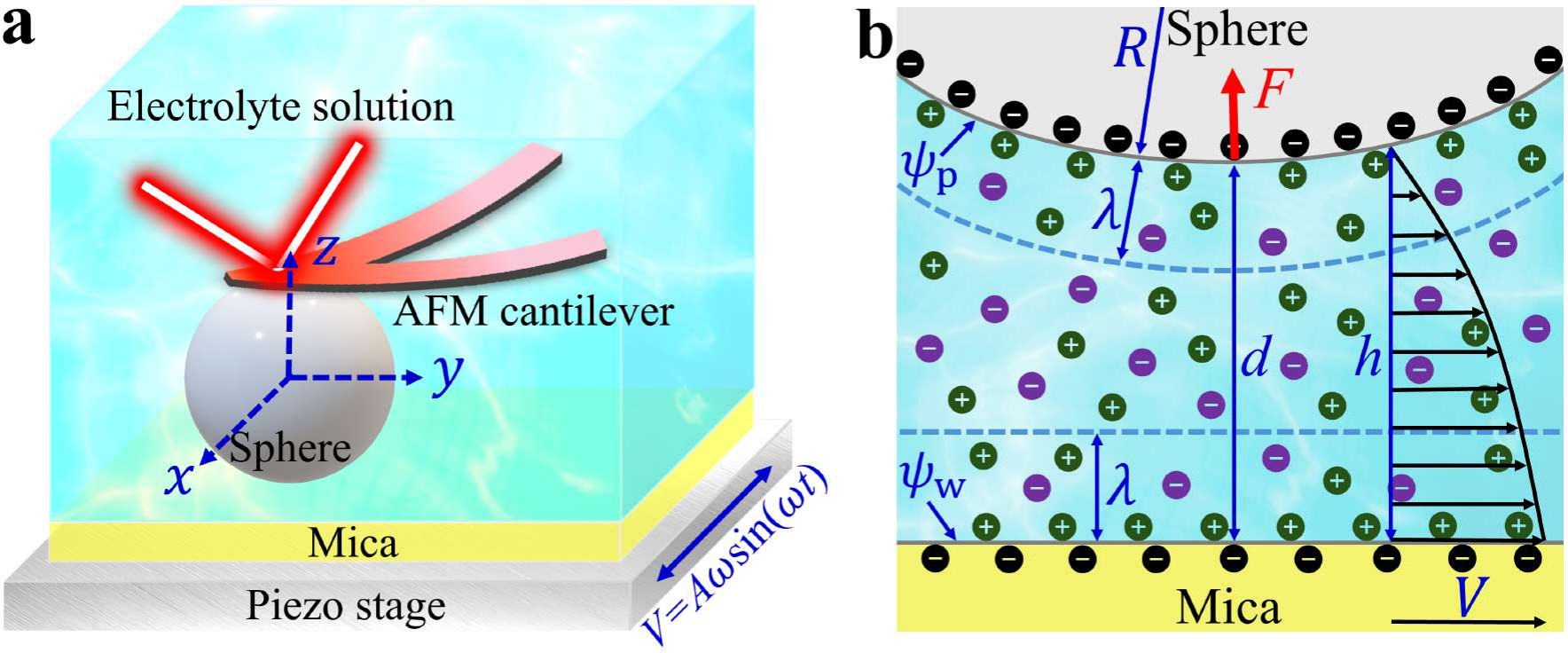}
\caption{ (a) Schematic illustration of the experimental setup. 
A borosilicate glass sphere is fixed to the tip of an AFM cantilever, which acts as a force sensor. The mica substrate, immersed in electrolyte solution, is mounted on a piezoelectric stage that imposes controlled  oscillations along the lateral $x$-direction. (b) Cut in the $(x,z)$ plane illustrating the ionic concentrations in the electric double-layers and the flow profile, and specifying the notations (see text).}
\label{Fig1}
\end{figure}

A sketch of the experimental configuration is shown in Fig. \ref{Fig1}. The experiment was performed using an AFM (Dimension 3100, Bruker) equipped with a liquid cell (DTFML-DD-HE). Borosilicate glass spheres (MO-Sci Corporation) with radii of $R$ = 56.6 $\pm$ 0.2 $\mu$m and $R$ = 24.4 $\pm$ 0.2 $\mu$m were glued to the ends of two AFM cantilevers (SNL-10, Bruker Probes). The stiffness of the cantilevers attached with spheres are calibrated using a hydrodynamic drainage method \cite{craig2001situ}, yielding values of $k_c=0.40\pm0.02$ N/m and $k_c=0.27\pm0.02$ N/m, respectively. The mica substrate was fixed on a piezo stage (E-709.CHG, PI), which enabled lateral oscillations  along the $x$-direction  at angular frequencies $\omega$ ranging from 126 to 503 s$^{-1}$, with amplitudes  $A$ between 7.2 and 17.1 $\mu$m. The imposed velocity is $V(t)=V_0\cos(\omega t)$, with probed velocity amplitudes $V_0=A\omega$ ranging between $1.3$ and $8.1$ mm/s. \textcolor{black}{The distance between the sphere and the mica surface is varied using the AFM piezo vertical displacement at a   velocity of 0.08 $\mu$m/s, which is sufficiently low to prevent any significant drainage force (see SM, fig. \ref{fig:erro}(c)).}

The electrolyte solution used in all experiments was deionized water containing Na$^+$ and Cl$^-$ at a concentration $c_\infty=0.1$ mM. This salt concentration was chosen so that electric effects are dominated by  Na$^+$ and Cl$^-$ ions (and not by carbonate ions due to the dissolution of CO$_2$). At the same time, with this value of $c_\infty$, the Debye screening length $\lambda =\sqrt{\epsilon k_\text{B}T/(2e^2c_\infty)}$ (with $\epsilon$ the liquid’s permittivity, $k_\text{B}T$  the thermal energy, $e$ the elementary charge) is long enough ($\sim 30$ nm) to enable the measurements of electric interactions.

\begin{figure}
\includegraphics[width=8cm]{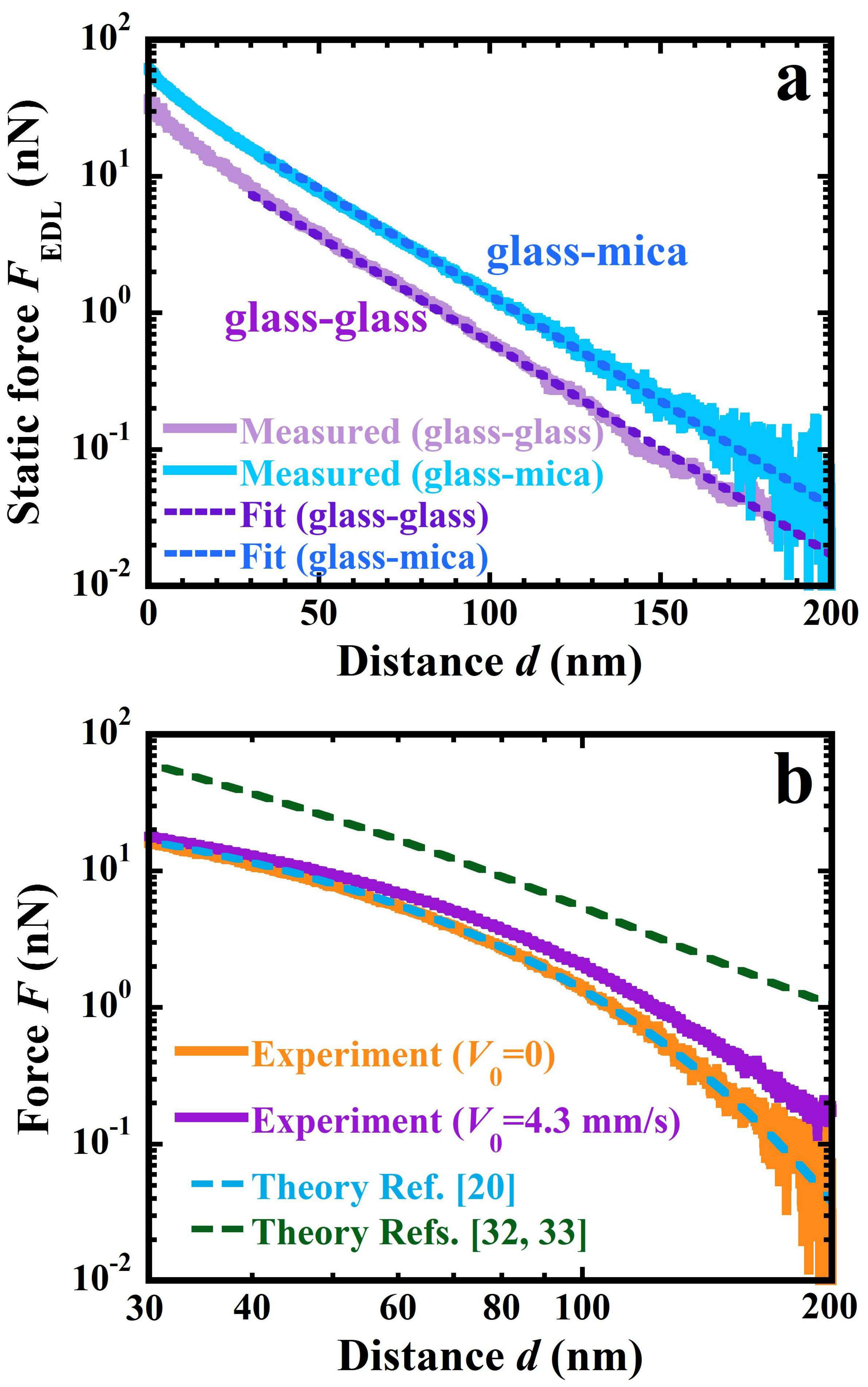}
\caption {(a) Measured  electric double-layer force ($F_{\text{EDL}}$) between a borosilicate glass sphere (radius: 56.6 $\mu$m) and either a borosilicate glass substrate or a mica substrate in 0.1 mM NaCl solution under static conditions (zero velocity).  Dashed curves represent theoretical fits based on Eq.~(\ref{Fstatic}), yielding $\lambda=28\pm 1$ nm, $\psi_\text{p}=-36 \pm 3$ mV, and $\psi_\text{w}=-100\pm5$ mV. (b) Log–log plot of the measured electric double-layer force (orange) and total time averaged force  (purple, sum of the electric double-layer and electroviscous lift forces), obtained with a 56.6 $\mu$m-radius sphere, and lateral  velocity amplitude  of $V_0=4.3$ mm/s. For comparison, total forces, as the sum of the static force Eq.~(\ref{Fstatic}) and the lift forces predicted by  Prieve and co-workers \cite{bike1990electrohydrodynamic} [Eq. (\ref{FPrieve}) in SM] or Warszynski \textit{et al.} \cite{warszynski2000electroviscous,warszynski1998electrokinetic} [Eq. (\ref{Fvs}) in SM], taking into account that $\langle V^2\rangle= V_0^2/2$, are also shown.}
\label{Fig2}
\end{figure} 

To determine the electric potential at the surface of the wall ($\psi_\text{w}$) and of the spherical particle ($\psi_\text{p}$), \textcolor{black}{the force in static conditions, called $F_{\text{EDL}}$, is first measured (Fig. \ref{Fig2}a). Its } theoretical expression reads \cite{alexander1987hydrodynamic,bike1995electrokinetic, wu1996electrokinetic,israelachvili2011intermolecular}:
\begin{align}
F_{\text{EDL}}=\frac{64\epsilon\pi R }{\lambda}\left(\frac{k_\text{B}T}{e}\right)^2\tanh\frac{e\psi_\text{p}}{4k_\text{B}T}\tanh\frac{e\psi_\text{w}}{4k_\text{B}T} e^{-d/\lambda} \label{Fstatic},
\end{align}
where the minimal distance $d$ between the particle and the wall is assumed to be larger than the Debye length. 
First, the static electric double-layer force between a borosilicate sphere and borosilicate glass substrate was measured as a function of
distance.  Fitting the measured force with Eq.~(\ref{Fstatic}), with $\psi_\text{w}=\psi_\text{p}$,  yielded a surface potential of $\psi_\text{p}=-36\pm3$ mV for the borosilicate sphere. Once this value was determined, the surface potential of the mica substrate was obtained by measuring the static force between the borosilicate sphere and the mica surface in the same solution, yielding $\psi_\text{w}=-100\pm5$ mV. 

Figure  \ref{Fig2}b shows the measured force as a function of the gap distance between the glass sphere and the mica surface. The force was measured for a lateral velocity amplitude of $V_0=4.3$ mm/s, and the static double-layer force, obtained without applying lateral motion, is also shown for comparison. At small separations, the two forces nearly coincide; however, at larger distances, the force measured during substrate motion exceeds that of the static case. 
This excess force cannot be explained by the inertial lift force which can be evaluated \cite{cherukat1994inertial,davies2018elastohydrodynamic} as $  1.7\rho V_0^2 R^2/2\sim 0.05$ nN, with $\rho$ the mass density of the fluid.
This observation provides direct evidence of the electroviscous lift force, which repels the sphere from the moving surface. Hereafter, we define the lift force $F_{\text{lift}}\equiv F-F_{\text{EDL}}$ as the difference between the total force $F$ and the static electric double-layer force $F_{\text{EDL}}$.

On the same graph, we represent %
existing predictions for this lift force available for the sphere wall geometry. The first one is the theory of Prieve and co-workers \textcolor{black}{\cite{bike1990electrohydrodynamic}} [see Eq.~(\ref{FPrieve}) in the Supplemental Material (SM)], where the lift force arises from the Maxwell stress associated to the electric field generated by the hydrodynamic flow imposed by the bead's lateral motion, in the lubrication approximation.  As can be seen in Fig. \ref{Fig2}b, this force cannot explain the increase of the total force induced by  the lateral oscillation of the sphere. 
This may not be surprising since it was already pointed out  \cite{cox1997electroviscous} that the electric field induced by the hydrodynamic flow generates an additional flow, associated to a viscous stress tensor that dominates over the Maxwell electric stress. 
A second prediction is that of Refs. \cite{warszynski2000electroviscous,warszynski1998electrokinetic} [see Eq.~(\ref{Fvs}) in SM], where the general formalism of Cox \cite{cox1997electroviscous} was applied to the cylinder-wall geometry in the lubrication regime  and generalized to the sphere-wall geometry using Derjaguin's approximation. As observed in Fig. \ref{Fig2}b, this lift force  is about one order of magnitude larger than the experimentally observed lift force. Another prediction for the sphere-wall geometry \cite{tabatabaei2006electroviscousSphere}, is not represented in Fig. \ref{Fig2}b since, for our parameters, it predicts an attractive lift force. 

 \begin{figure*}
\includegraphics[width=18cm]{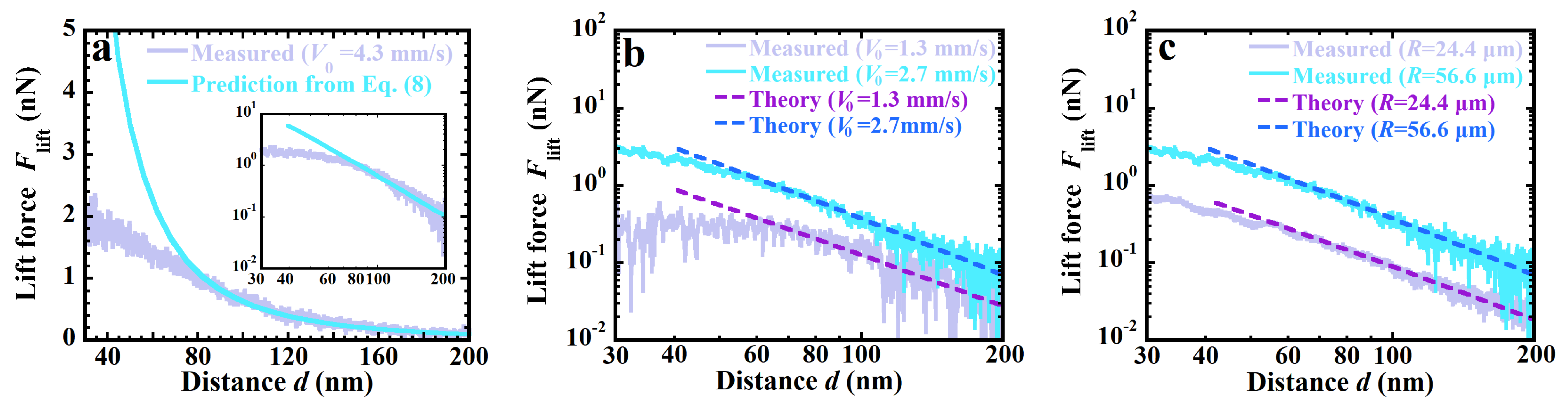}
\caption{ (a) Lift force ($F_{\text{lift}}$), obtained by subtracting the static double-layer force ($F_{\text{DL}}$, measured at zero velocity) from the total force ($F$) measured during lateral oscillations, for a velocity of $V_0=4.3$ mm/s and a sphere radius $R=56.6$ $\mu$m, as a function of separation distance $d$. The continuous light blue line represents the prediction of Eq. (\ref{ScalingPredictionLast}). Inset: same data on a logarithmic scale. 
(b) Lift forces measured at two velocities, $1.3 $ mm/s and $2.7$ mm/s, with a sphere radius of $R=56.6$ $\mu$m. Dash-dotted lines are the theoretical predictions. 
(c) Measurements performed with spheres of radii 56.6 $\mu$m and 24.4 $\mu$m at $V_0=2.7$ mm/s.  
 }
\label{Fig3}
\end{figure*} 

We now present a theoretical analysis of the problem, in the case that the sphere is fixed and the wall moves at constant velocity $V$ (\textcolor{black}{we show later how to adapt this analysis to time-dependent velocities)}. 
The equations to be solved are the transport equation for the diffusion of ions in the presence of hydrodynamic flow and electric field,  the Poisson equation for the electric potential generated by the ions, and the Stokes equation for the flow (which also depends on the concentration of ions). Since all phenomena (ion transport - flow - electric field) are coupled in non-linear ways, this is a challenging theoretical problem. When the distance between the particle and the wall is large compared to the screening length $\lambda$, analytical progress can be made by using matched asymptotic expansion, separating the medium into inner layers (at distances of the order of the Debye length from the walls) and an outer region (far from the walls), where electroneutrality is achieved, but where effective boundary conditions describing the exchange of ions with the electric double-layer. This was done by Cox~\cite{cox1997electroviscous}. Here, we analyze these equations by adapting the standard lubrication approach. Clearly, $d$ is the relevant lengthscale for the vertical direction, while in the lateral directions the relevant lengthscale is $l=\sqrt{Rd}$, since the local height can be written as $H=h/d=1+(X^2+Y^2)/2$, where $X=x/l$ and $Y=y/l$ are rescaled horizontal coordinates. A relevant parameter is the Péclet number $\text{Pe}=Vl/D_e$ associated to this lengthscale $l$, where $D_e=2D_+D_-/(D_++D_-)$ is the effective diffusion coefficient of ions. Here,  $D_+$ and $D_-$ are the respective diffusivities of cations and anions, which for NaCl read $D_+=1.33 \times 10^{-9}\, \mathrm{m}^2/\mathrm{s}$ and $D_-=2.03 \times 10^{-9}\, \mathrm{m}^2/\mathrm{s}$. 

We first consider the concentration of cations $c$  at distances larger than the Debye length from the interfaces. Since electroneutrality holds in this outer region, $c$ is also the concentration of anions. Applying the lubrication approach  to Cox equations, we find
an effective  two-dimensional advection diffusion  equation for the rescaled concentration  $N= (c-c_\infty) d^2/ (c_\infty \lambda^2)$ that reads (see SM for details) 
\begin{align}
\text{Pe}\   \overline{\ve[U]}_\parallel \cdot  \nabla_\parallel N &=    \nabla_\parallel\cdot(H  \nabla_\parallel N) + 2  \text{Pe} (\alpha_\text{w} S_\perp^\text{w}    +\alpha_\text{p} S_\perp^\text{p} )   \label{AdvDiffSource},
\end{align}
where $\nabla_\parallel=\hat{\ve[e]}_x\partial_X+\hat{\ve[e]}_y\partial_Y$,  $\overline{\ve[U]}_\parallel $  is the  flow \textcolor{black}{parallel to the wall}  integrated over the vertical coordinate \textcolor{black}{(rescaled by $Vd$)}, and the dimensionless parameters $\alpha_\text{s}$ are defined by
\begin{align}
\alpha_\mathrm{s}= \frac{\psi_\text{s}\ e}{k_\text{B}T}\frac{D_+-D_-}{D_++D_-}+4\ln\cosh \frac{\psi_\text{s}\ e}{4k_\text{B}T}, \ \ s\in\{w,p\}.
\end{align}
Finally, the source term $S_\perp^\text{s}$ is proportional to the normal component of the flow to the surface $s$ and represents the exchange of ions with the Debye layer. Importantly, the effective flow and source terms can be derived explicitly using the known value of the flow induced by the lateral motion of the wall \cite{oneill1967slow}, leading to 
\begin{align}
&\overline{\ve[U]}_\parallel = \left(\frac{3}{5} + \frac{X^2+3Y^2}{10}\right)\hat{\ve[e]}_x -\frac{ X Y}{5}\hat{\ve[e]}_y \label{ValueU},\\
&S_\perp^\text{s} = \frac{1}{10H^3}\times \begin{cases}
  X \left(26+X^2+Y^2\right) & (\text{s}=\text{p})\\
   4 X (4 -  X^2 - Y^2)  & (\text{s}=\text{w}) \label{ValueSource}
   \end{cases}.
\end{align} 

 Next, Cox theory takes into account the flow that is generated by the non-equilibrium concentration of ions, the lift force is dominated by the associated viscous stress. We show in the SM that, in the lubrication regime for a sphere wall geometry, this force  reads 
\begin{align}
&F_\text{lift}=F_0 \ \Phi(\text{Pe},\alpha_\mathrm{p},\alpha_\mathrm{w}),  &F_0 = \epsilon \left(\frac{k_\text{B}T\lambda}{e}\right)^2 \frac{R}{d^3},  \label{ScalingPrediction}
\end{align}
where  $F_0$ is a characteristic force and the dimensionless lift force $\Phi$,  depending only on the Péclet number and the parameters $\alpha_\mathrm{w},\alpha_\mathrm{p}$, reads
\begin{align}
\Phi= 3 (\alpha_\mathrm{p}+\alpha_\mathrm{w})\iint_{\mathcal{R}^2 } dX dY \ \frac{\nabla_\parallel N \cdot \nabla_\parallel H}{H^2}.  \label{Force1}
\end{align}
As seen in the above equation, the force is generated by the ionic concentration gradients. For a fixed value of the Péclet, it is straightforward to integrate the two-dimensional partial differential equations (\ref{AdvDiffSource}), (\ref{ValueU}), and (\ref{ValueSource}) for $N$, leading to the determination of $\Phi$. 
Note that the relative variation of ionic concentration due to electroviscous effects, in the bulk and in the  EDLs, is of the order $(\lambda/d)^2\ll1$. This means that the local Debye length is only very slightly changed with respect to its static value, and justifies the assumption that the static force, coming from the interactions between static EDLs, \textcolor{black}{is still present and } can be added to the (dynamic) lift force to obtain the total force, as commonly assumed \cite{wu1996electrokinetic,warszynski1998electrokinetic}.
Our theory is valid in the regime $\lambda\ll d\ll R$. 

To compare with the experiments, we take into account the fact that the velocity is not constant but reads $V(t)=V_0 \cos(\omega t)$. The relevant Péclet number is thus $\text{Pe}_*=V_0 l/D_e$. The prediction of the theory for the time-averaged force in this case is:
\begin{align}
&F_\text{lift} = F_0\  {\Phi}^*(\text{Pe}_*),  \label{ScalingPredictionLast}
\end{align}
where ${\Phi}^*(\text{Pe}_*)=\int_0^{2\pi}d\tau \ \Phi(\text{Pe}_*\cos\tau)/(2\pi)$.  Since the model is derived under the assumption of quasistatic velocities, we verified numerically (see \textcolor{black}{Fig. \ref{FigSafeZone} in} SM) that the values of $\omega$ used in the experiments are sufficiently small for this approximation to remain valid.

Analytical calculations can be performed in the small and large Péclet limits, respectively. For small Péclet, we obtain 
\begin{align}
\Phi^*  \underset{\text{Pe}_*\ll 1}{\simeq} \text{Pe}_*^2\frac{(\alpha_\text{p}+\alpha_\text{w})}{2}\left[  \frac{6\pi}{25}(\alpha_\text{w}+\alpha_\text{p})  + k(\alpha_\text{p}-\alpha_\text{w}) \right] \label{FSMallPéclet}
\end{align}
with $k\simeq 0.49$.  In this regime, the force is quadratic in the velocity, as in previous studies, and in particular the only difference with the result by Ref. \cite{tabatabaei2006electroviscousSphere} is the value of the coefficient $k$, which reads $k=6.26$ in their study. The origin of this discrepancy is explained in the SM. In the opposite limit of large Péclet,   we find
 \begin{align}
\Phi^* \underset{\text{Pe}_*\gg 1}{\simeq}   (\alpha_\text{p}+\alpha_\text{w}) \pi \left[(52-45\ln3) \alpha_\text{w}+\frac{64 -45\ln3}{2} \alpha_\text{p} \right]\label{FLargePéclet}
\end{align}
This means that, for large Péclet, the force no-longer grows quadratically with the velocity but saturates to a finite value, that no longer depends on the velocity. The origin of this saturation can be traced back to Eq.~(\ref{AdvDiffSource}), where both the advective term $\overline{\ve[U]}_\parallel\cdot\nabla N$ and the source term $S_\perp$ are proportional to the imposed velocity. As a consequence, even if increasing the velocity increases the flux of ions escaping from the Debye layer due to the flow, it also increases the velocity in the advective component, and the resulting concentration gradients (and thus, the lift force) become independent of the imposed velocity when this one is large. To the best of our knowledge, this saturation of the force at high Péclet has not been identified before.

Figure \ref{Fig3}a presents the results for the pure electroviscous lift force, obtained by subtracting the static double-layer force from the total force measured during lateral motion. The \textcolor{black}{continuous} line corresponds to the theoretical calculation based on our model [Eq. (\ref{ScalingPredictionLast})]. The experimental data show good agreement with the theoretical predictions for distances larger than the Debye screening length. At shorter distances, a saturation of the force is observed, which is not captured by our model, as it is only valid in the regime of large separations. Figure \ref{Fig3}b shows the measurements of the lift force obtained at two different velocities. It is clearly seen that the force increases with increasing velocity. Figure \ref{Fig3}c depicts the results obtained at the same velocity but using two spheres of different radii, it shows that the larger the sphere, the greater the resulting force.

\begin{figure}
\includegraphics[width=8cm]{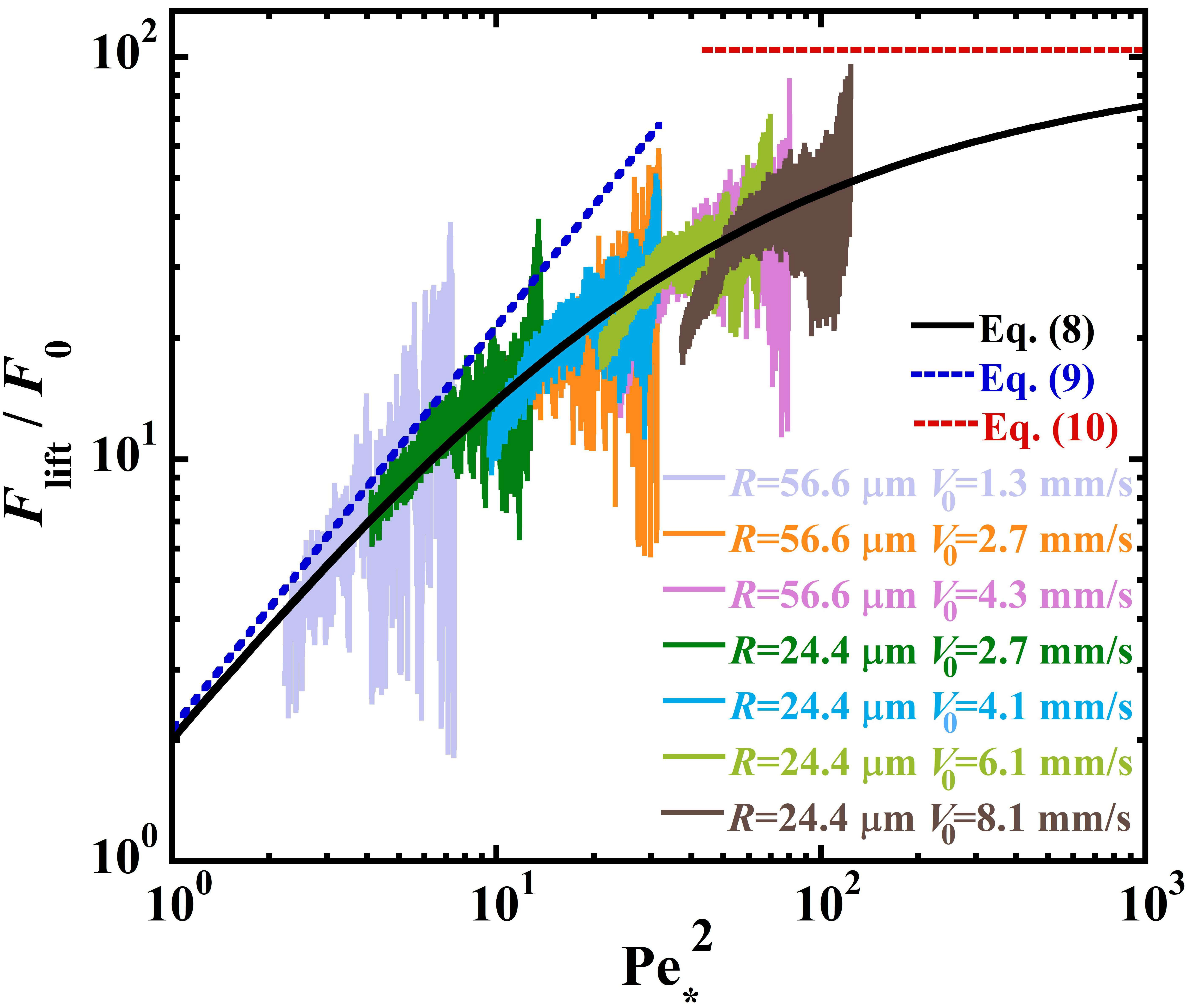}
\caption{  Dimensionless lift force $F_\text{lift}/F_0$ as a function of $\text{Pe}_*^2=V_0^2 Rd/D_e^2$. Experimental data obtained under various conditions are represented by different colors, for a range of $d$ between 60 and 200 nm. The black line denotes the theoretical electroviscous lift force calculated from Eq.~(\ref{ScalingPredictionLast}). Dash-dotted blue and red lines represent respectively the predictions at low [Eq.~(\ref{FSMallPéclet})] and large [Eq.~(\ref{FLargePéclet})] Péclet numbers.
}
\label{Fig4}
\end{figure} 

To rationalize our measurements, the experimentally measured electroviscous lift force was normalized by $F_0$ and plotted as a master curve as a function of the Péclet number, in accordance with the theoretical prediction Eq. (\ref{ScalingPredictionLast}). 
Figure \ref{Fig4} shows the measurements obtained for various velocities and sphere radii. We excluded distances $d<2\lambda\simeq 60$ nm (where the theory is not valid) and distances larger than $200$ nm (where the noise is too high).
The results follow the theoretical prediction of our model: as the Péclet number increases, the normalized force rises and tends to a saturation. In our experiments, however, we could not reach the saturation limit, because very large Péclet numbers correspond to large separations, where the measured force becomes extremely small and the signal-to-noise ratio consequently decreases. Nevertheless, our data unambiguously show a deviation of the rescaled force from a quadratic behavior with Péclet number.

In conclusion, we have presented measurements of the electroviscous lift force acting on an electrically charged sphere sliding along a charged solid surface in an aqueous solution. The lift force was investigated as a function of the separation distance between the sphere and the substrate, for various sliding velocities and two distinct sphere radii. Experimental results were compared with existing theoretical models for the electroviscous force; however, none of these models were able to accurately describe our observations. To address this discrepancy, we derived a new analytical expression for the lift force based on Cox’s formalism within the lubrication approximation framework,which predicts a new regime where the electroviscous force saturates for large driving velocity. Our experimental data show excellent quantitative agreement with this new theoretical prediction, in the absence of any fitting parameter. Altogether, our results provide the first robust and quantitative measurements of the electroviscous lift force that emerge between charged particles as a result of the motion of ions in viscous electrolytes in non-equilibrium conditions. 

\begin{acknowledgments}

Acknowledgments - The authors thank Alois Würger for fruitful discussions. H. Zhang and Z. Zhang acknowledge the financial support of the China Scholarship Council during their stay in Bordeaux, which enabled them to carry out the experiments. H. Zhang acknowledges support from the China Postdoctoral Science Foundation (Grant No. 2026M793764). Z. Zhang acknowledges support from the Beijing Natural Science Foundation (Grant No. 2262016) and the National Natural Science Foundation of China (Grant No. 12404234). The authors thank the French National Research Agency for the supporting grants MS-DOS (Grant No. ANR-25-CE06-3953-02) and EDDL (Grant No. ANR-19-CE30-0012). 
\end{acknowledgments}

Author contributions - A.M. conceived the project. T.G. developed the theory. H.Z., Z.Z. and A.M. performed the experiment and analysed the data. H.Z., A.M. and T.G. drafted the manuscript. All authors contributed to the interpretation of the results and the final version of the manuscript.

\appendix

\vspace{4cm}
\newpage
\begin{center}
{\bf SUPPLEMENTAL MATERIAL }
\end{center}

\vspace{1cm}
Here, we provide the following elements:

\begin{itemize}
\item We show additional experimental information (Section \ref{Secadditional}),
\item We write the general equations to be solved, their simplification in Cox's theory, and we review  existing results for the lift force (Section \ref{SecEq}),
\item We show how to simplify these equations in the lubrication regime (Section \ref{CoxLubrication}),
\item We give explicit solutions of the equations in the regimes of small and large Péclet numbers,  with a comparison with previous results when they exist (Section  \ref{SecExplicitSol3D}),
\item  In Section \ref{SectionControls}, we control the validity of our hypotheses, by checking that non-stationary and inertial effects can be neglected in our study,  and by commenting on the validity of Cox's equations.
\end{itemize}

\section{Additional data for the experiments}
\label{Secadditional}
\subsection{Calibration the AFM cantilever stiffness}

The spring constants of the colloidal probes were calibrated using the drainage-force method described in Ref.~\cite{craig2001situ}. This technique is based on the hydrodynamic resistance that develops when an AFM probe approaches a rigid substrate in a liquid environment. By measuring the resulting hydrodynamic force ($F_{\text{D}}$) during the vertical motion of the cantilever relative to the substrate, the cantilever stiffness ($k_{\text{c}}$) can be accurately determined.

The hydrodynamic drainage force acting on the sphere is given by
\begin{equation}
F_D = -\frac{6\pi \eta R^2 \dot{d}}{d},
\label{eq:S1}
\end{equation}
where $\eta$ is the dynamic viscosity of the liquid, $R$ is the sphere radius, $d$ is the sphere--surface separation distance, and $\dot{d}=\partial_t d$ is the relative approach velocity that we obtain from the time derivative of the gap distance. The cantilever deflection ($Z_{\text{c}}$) is related to the force by
\begin{equation}
F_D= k_c Z_c.
\label{eq:S2}
\end{equation}
Using Eq.~(\ref{eq:S1}) and Eq.~(\ref{eq:S2}), we obtain:
\begin{equation}
\vert\frac{Z_c}{\dot{d}}\vert = \frac{6\pi \eta R^2}{k_c d}.
\label{eq:S3}
\end{equation}

\begin{figure}[ht!]
    \centering
    \includegraphics[width=8cm]{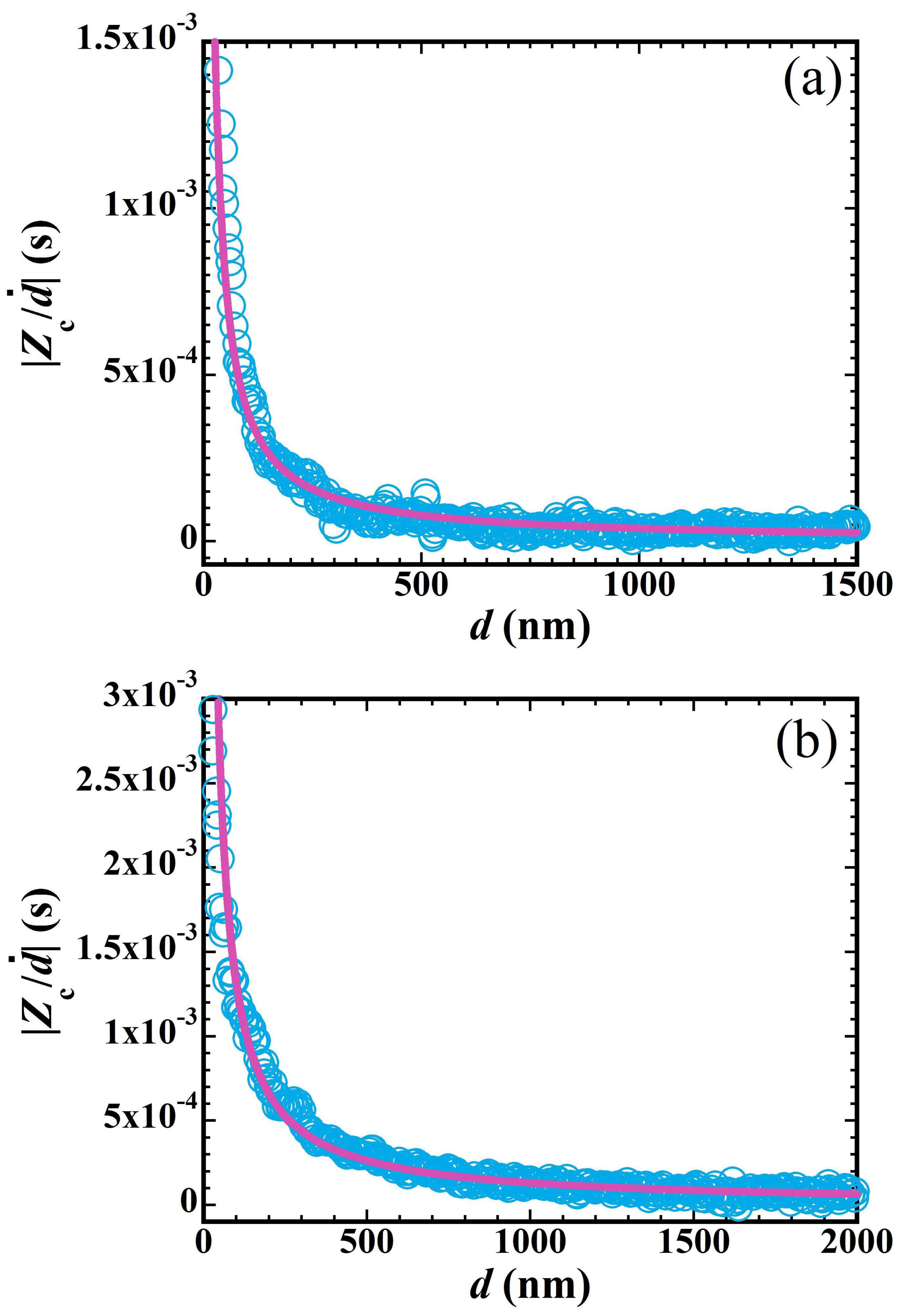}
    \caption{
Measured $\vert Z_c/\dot{d}\vert $ versus separation distance at 100 mM NaCl solution to minimize the contribution of double-layer force. The solid lines show theoretical fits based on Eq.~(\ref{eq:S3}), yielding cantilever stiffness of $0.27 \pm 0.02~\mathrm{N/m}$ and $0.40 \pm 0.02~\mathrm{N/m}$ for sphere radii of $24.4~\mu\mathrm{m}$ (a) and $56.6~\mu\mathrm{m}$ (b), respectively.
    }
    \label{fig:Calibration}
\end{figure}

The cantilever stiffness is determined by fitting the experimentally measured $\vert Z_c/\dot{d}\vert$ as a function of the separation distance $d$ using Eq.~(\ref{eq:S3}), while keeping $\eta$ and $R$ fixed to their known values. The resulting fit, together with the experimental measurements, is presented in Fig.~\ref{fig:Calibration}.

\begin{figure*}
    \centering
    \includegraphics[width=16cm]{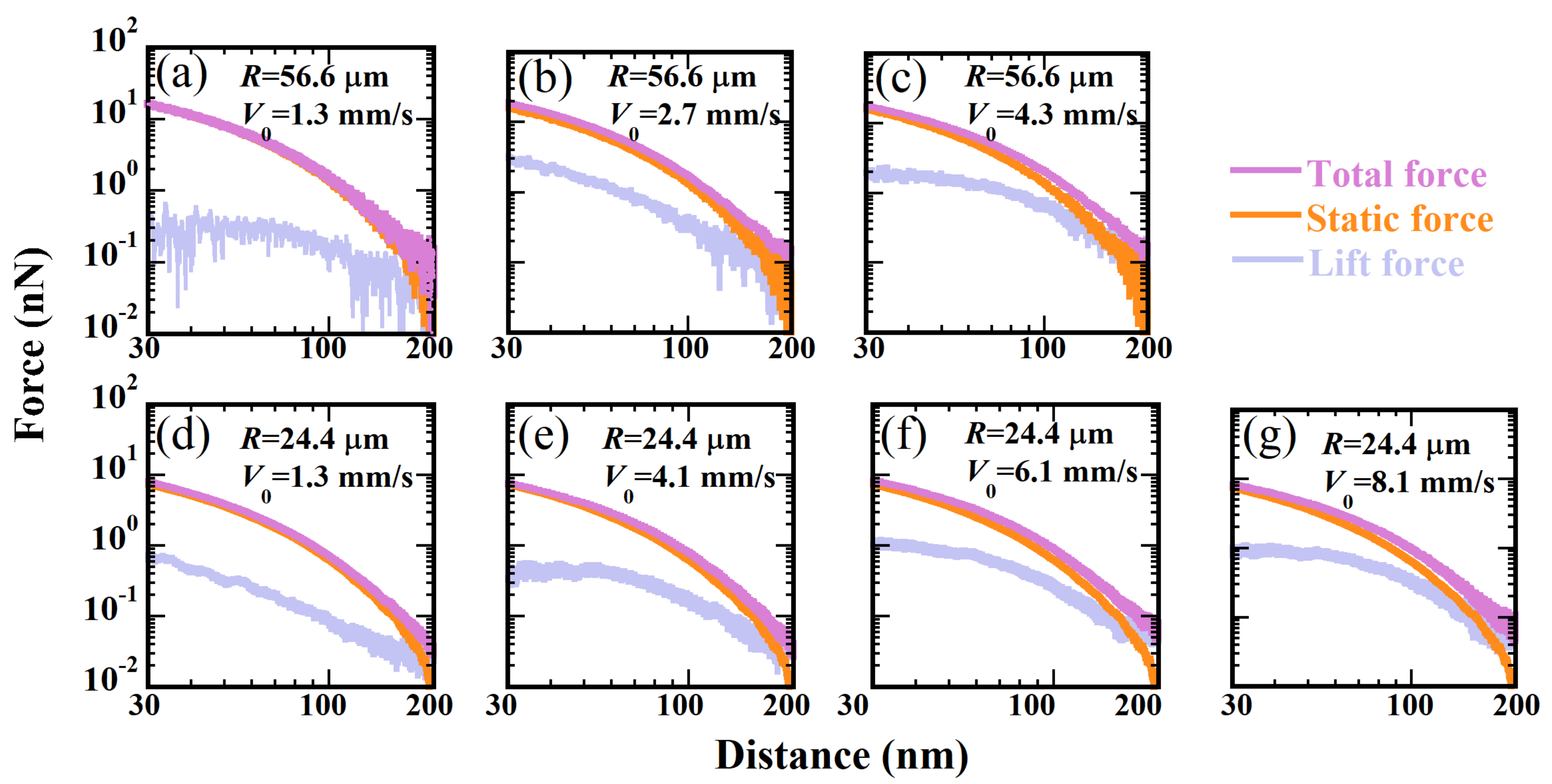}
\caption{ Log–log plot of the measured static electric double-layer force (orange, without lateral velocity, $V_0=0$) and time-averaged force obtained under lateral oscillation of the mica substrate (purple, total force: sum of the electric double-layer and electroviscous lift forces) as a function of separation distance under different experimental conditions. The lift force is determined by subtracting the electric double-layer force from the measured total force.}
    \label{fig:alldata}
\end{figure*}

\subsection{ Additional experimental data}
In Fig.~\ref{fig:alldata}, we present the force–distance measurements recorded for different sphere radii $R$ and lateral velocity amplitudes $V_0$. For each experimental condition, the measured time-averaged force (total force) includes both electroviscous lift force and electrostatic double-layer force. The lift force is then obtained by subtracting the measured static double-layer force from the measured total force. \textcolor{black}{In the data analysis, the measured force curves were first smoothed, and we set the force at large separations to zero before performing the data subtraction. The data obtained with and without lateral motion were acquired in the same series of experiments.}

\subsection{ Discussion on experimental errors}
\begin{figure*}
    \centering
    \includegraphics[width=18cm]{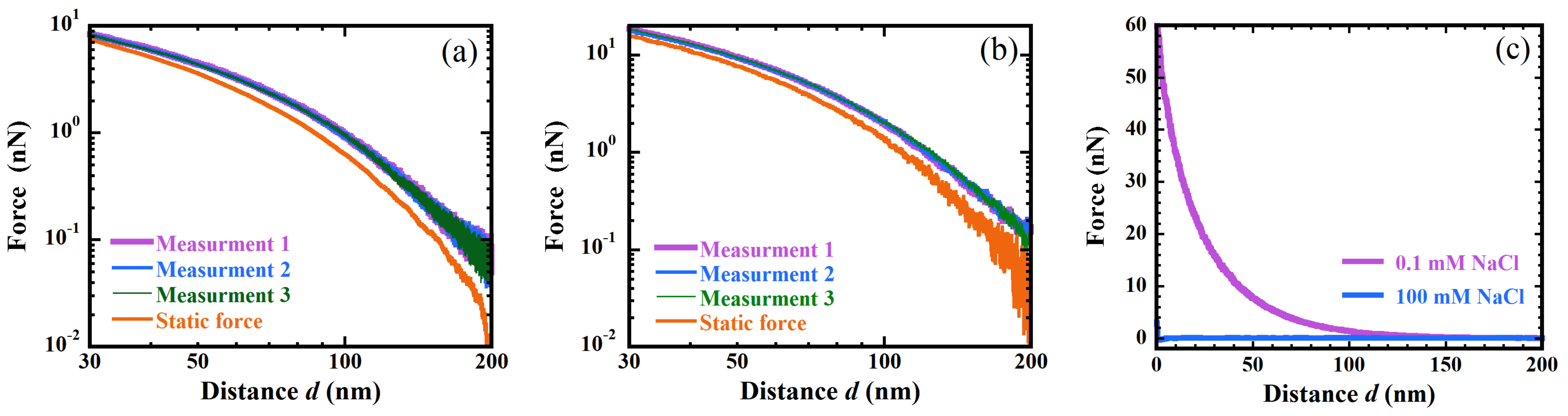}
    \caption{ \textcolor{black}{Examples of three independent measurements with spheres of radius 24.4 $\mu$m at 8.1 mm/s (a) and 56.6 $\mu$m at 4.3 mm/s (b), corresponding to Fig S2g and S2c, respectively. (c) Force measurements performed in 0.1 and 100 mM NaCl solutions with a sphere of radius 56.6 $\mu$m, without lateral motion. No measurable force is observed in the 100 mM NaCl solution (where the electrostatic double-layer force was screened), indicating that the contribution of the drainage force is negligible under our experimental conditions.}
        }
    \label{fig:erro}
\end{figure*}

 The uncertainty in the experimentally measured force primarily arises from the uncertainty in the cantilever stiffness. Since the force is calculated as the product of the AFM cantilever deflection and the stiffness, the overall uncertainty in the force measurement is estimated to be less than $8\%$. This uncertainty is not represented on the figures because it is smaller than experimental scatter. \textcolor{black}{The measurements exhibit good reproducibility, as evidenced by the negligible variation among the three independent measurements shown in Fig.~\ref{fig:erro}a and b. Another potential source of error in the experiment is the drainage force arising as the sphere approaches the surface. To minimize this effect, the approach velocity in the normal direction was controlled at a low value of $0.08\ \mu$m/s. Under this condition, the drainage force is below the experimental detection limit, as demonstrated in Fig.~\ref{fig:erro}c}

 When comparing with the theory, there is an uncertainty on the theoretical prediction due to the uncertainties on the sphere radius, the velocity and the surface potentials. The error on the surface potentials is the largest. Estimating the uncertainty on the predicted force with the formula of error propagation, we find that the prediction for the force is plus or minus $11\%$, which is also within the experimental scatter of the data.

\section{Problem statement}
\label{SecEq}
\subsection{Physical situation}

We consider a  fixed charged sphere of radius $R$, close to a flat wall moving at a constant velocity $V$ in the $x$ direction. The vertical $z$ direction is perpendicular to the flat lower wall, and in 3D we call $y$ the third spatial direction, so that $(x,y,z)$ are Cartesian coordinates. 

In the stationary state, the concentrations of cations ($c_+)$ and anions ($c_-$), the hydrodynamic flow $\ve[v]$, and the electric potential $\psi$ are governed by the following equations,  for a symmetric electrolyte:
\begin{align}
&\boldsymbol{\nabla}\cdot\ve[v]=0, \ \ \ 
-\boldsymbol{\nabla} p+\eta\boldsymbol{\nabla}^2 \ve[v]=q(c_+-c_-)\boldsymbol{\nabla}\psi, \label{4312}\\
&\boldsymbol{\nabla}^2\psi=-q (c_+-c_-)/\epsilon,\\
&\ve[v]\cdot\boldsymbol{\nabla} c_+=D_+\boldsymbol{\nabla}\cdot \left(   c_+\frac{q  \boldsymbol{\nabla}\psi }{k_\mathrm{B}T}+\boldsymbol{\nabla} c_+  \right), \label{AdvDiffPlus}\\
&\ve[v]\cdot\boldsymbol{\nabla} c_-=D_-\boldsymbol{\nabla} \cdot\left( -c_-\frac{q  \boldsymbol{\boldsymbol{\nabla}}\psi  }{k_\mathrm{B}T}+\boldsymbol{\boldsymbol{\nabla}} c_-  \right), \label{AdvDiffMinus}
\end{align}
Here, $q>0$ is the charge of a single cation, each anion carries the opposite charge $-q$.  Note that $q=\tilde{Z}e$, with $\tilde{Z}$ the valency of the cation, the experimental situation corresponds to $\tilde{Z}=1$. Next,  $T$ is the temperature, $k_\mathrm{B}$ is Boltzmann's constant, $\eta$ is the viscosity, $\epsilon$ is the permittivity of the liquid, $p$ the pressure, and $D_\pm$ respectively represent the diffusivity of cations and anions. 

At the lower wall, the flow velocity is the same as that of the wall, $\ve[v]=V\hat{\ve[e]}_x$, where $\hat{\ve[e]}_i$ represents the unit vector in the direction $i\in\{x,y,z\}$. The velocity $\ve[v]$ vanishes at the surface of the sphere.  For the electric potential we consider constant potential conditions, with $\psi=\psi_\text{w}$ at the lower wall (w) and $\psi=\psi_\text{p}$ at the surface of the particle (p). 
Last, we have no flux conditions for ions at the solid surfaces
\begin{align}
&\ve[n]\cdot \left( (\pm c_\pm)\frac{q   \boldsymbol{\nabla}\psi }{k_\mathrm{B}T} +{\boldsymbol{\nabla}} c_\pm  \right)=0, & (\ve[r]\in\partial\Omega_\text{s}),
\end{align}
where $\ve[n]$ is the normal vector to the solid surfaces, which point towards the liquid by convention. Note that $\partial\Omega_\text{s}$, with $\text{s}\in\{\text{w},\text{p}\}$, denotes either the surface of the lower flat wall (s$=$w) or the spherical particle's surface (s$=$p). We also introduce $d$ as the minimal distance between the sphere and the wall. Finally, far from the boundaries, the potential $\psi$ vanishes, and the concentrations $c_\pm$ tend to the value the concentration of ions $c_\infty$ in the solution. 
 
Our goal is to calculate the force on the sphere
\begin{align}
\textcolor{black}{F} = \int_{\partial\Omega_p} dS \  \ve[n]\cdot\ve[\Sigma] \cdot\hat{\ve[e]}_z,
\end{align}
where    the stress tensor $\ve[\Sigma]$ is the sum of the viscous stress and the Maxwell electromagnetic stress:
\begin{align}
\ve[\Sigma] =-p  \ve[I] + \eta[{\boldsymbol{\nabla}}\ve[v]+({\boldsymbol{\nabla}}\ve[v])^\dagger] +\epsilon \left({\ve[E]}\otimes{\ve[E]}-\frac{ {\ve[E]}^2}{2} \ve[I]\right), \label{StressGen}
\end{align}
with ${\boldsymbol{\nabla}}\ve[v]$ the gradient of $\ve[v]$, $\ve[I]$ the identity tensor and $ {\ve[E]}=-{\boldsymbol{\nabla}}\psi$ the electric field.   
We also introduce the Debye screening length as
\begin{align}
 \lambda=\sqrt{\frac{k_\mathrm{B}T\epsilon}{2q^2 c_\infty}}. 
\end{align}

\subsection{Existing results for the sphere wall geometry}

We report   existing results for the lift force  for the sphere-wall geometry. We first consider the theory of  Prieve and co-workers, who calculated the Maxwell stress (MS) on the spherical particle in the regime $ \lambda \ll d \ll R $, giving rise to the force \cite{bike1990electrohydrodynamic}
\begin{align}
&F_\text{MS} = \frac{\epsilon^3 \pi V^2 R}{K^2 d^3} 
\Bigg[ 0.384 \left( \frac{\psi_p + \psi_w}{2} \right)^2+ \nonumber\\
&\quad 0.181 \frac{(\psi_p - \psi_w)(\psi_p + \psi_w)}{2} 
+ 0.024 (\psi_p - \psi_w)^2 \Bigg] \label{FPrieve}
\end{align}
  where $K=\epsilon(D_++D_-)/(2\lambda^2)$ \cite{bike1990electrohydrodynamic} is the conductivity of the liquid. As can be seen in Fig. 2b of the main text, this force cannot explain the increase of lift force due to the lateral oscillation of the sphere. This may not be surprising since it was already pointed out \cite{cox1997electroviscous} that the electric field induced by the hydrodynamic flow generates an additional flow, associated to a viscous stress tensor that dominates over the Maxwell electric stress. In this reference, a general formalism was introduced to calculate this viscous stress in the limit of a small Debye screening length. 
  Using this formalism in the cylinder-wall geometry,  under the lubrication approximation, for small enough velocities and assuming  large distances ($d\gg\lambda$), the expression of the force due to the electro-viscous stress was calculated. By further applying the Derjaguin approximation, Warszynski \textit{et al.} \cite{warszynski2000electroviscous,warszynski1998electrokinetic} obtained the electroviscous lift force acting on the sphere from cylinder-wall geometry  as
\begin{align}
&F_{\text{VS}} = \frac{\epsilon\pi\lambda^2}{D_e^2} \left( \frac{k_\text{B}T}{q} \right)^2 \frac{V^2 R^2}{d^2}(D_+ I_{w}^- + D_- I_{w}^+) \nonumber\\
&\times  \frac{(9D_+ I_{p}^- + D_+ I_{w}^- + 9D_- I_{p}^+ + D_- I_{w}^+)}{6 (D_+ + D_-)^2}  \nonumber\\
&+ \frac{\epsilon \pi\lambda^4}{D_e^2} \left( \frac{k_\text{B}T}{q} \right)^2 \frac{V^2 R}{d^3} \nonumber\\
&\times \Bigg[ \frac{ 7 (D_+ I_{p}^- - D_- I_{p}^+)^2 + (D_+ I_{w}^- - D_- I_{w}^+)^2 }{30 (D_+ + D_-)^2} \nonumber\\
&\quad + \frac{ 4 D_+ D_- \left( \frac{D_+}{D_-} I_{p}^- I_{w}^+ + \frac{D_-}{D_+} I_{p}^+ I_{w}^- - I_{p}^+ I_{w}^- - I_{p}^- I_{w}^+ \right) }{30 (D_+ + D_-)^2} \Bigg] \label{Fvs},
\end{align}
where  $I_{\text{s}}^{\pm} =  \mp \Psi_{s} + 4 \ln  \cosh \left( \frac{\Psi_s}{4} \right) $, $\Psi_s={\psi}_s q/k_\text{B}T$, and $D_e=\frac{2D_+D_-}{D_++D_-}$ denotes the effective diffusion coefficient. The first term of Eq. (\ref{Fvs}) arises from viscous stress and the second term is due to the Maxwell stress. Within the range of distances for which this expression is valid ($ \lambda \ll d$), the second term is much smaller than the first one, with their ratio scaling as $\lambda^2/Rd$. As observed in Fig. 2b in the main text, this force is about one order of magnitude larger than the experimentally observed lift force.   

Finally, Tabatabaei \textit{et al.} used the Lorentz reciprocal theorem to calculate the  electroviscous lift force acting on a cylinder \cite{tabatabaei2006electroviscousCylinder} and a sphere \cite{tabatabaei2006electroviscousSphere} sliding along a planar wall at small velocity in the lubrication regime. 
In the sphere-wall geometry, they obtained
\begin{align}
&F_{\text{VS}}=\frac{24\pi\epsilon}{5}\left(\frac{k_\text{B}T\lambda V R}{q d}\right)^2\times\nonumber\\
&\Bigg\{\frac{1}{5}\left(\frac{G_\text{p}}{D_+}+\frac{H_\text{p}}{D_-}+\frac{G_\text{w}}{D_+}+\frac{H_\text{w}}{D_-}\right)^2\nonumber\\
&+1.667 \left[\left(\frac{G_\text{p}}{D_+}+\frac{H_\text{p}}{D_-}\right)^2-\left(\frac{G_\text{w}}{D_+}+\frac{H_\text{w}}{D_-}\right)^2 \right] \Bigg\},\label{TabaSphere}
\end{align}
see Eq.~(6.27) in Ref. \cite{tabatabaei2006electroviscousSphere}. Here the dimensionless parameters $(G_\text{p},H_\text{p})$ on the sphere and $(G_\text{w},H_\text{w})$ on the substrates are defined as 
\begin{align}
&G_\text{s}=\ln\frac{1+e^{-\Psi_s/2}}{2}, &H_\text{s}=\ln\frac{1+e^{\Psi_s/2}}{2}.
\end{align}
Note that, in the case where the sphere and the wall display the same electric properties, Eq. (\ref{TabaSphere}) was obtained independently in Ref. \cite{schnitzer2016streaming}. For the parameters corresponding to our experiments, the lift force Eq.~(\ref{TabaSphere}) is negative, which disagrees with our observations. We show in Section \ref{SecSmallV3DGen} that, although the above formula is exact for small velocities when the zeta potentials of the particle and the wall are identical, it is not valid for distinct zeta potentials due to a subtle mathematical issue of uniform convergence. 
 
\subsection{The limit of small Debye length: Cox's Equations}
\label{CoxEquations}
In the limit $\lambda\to0$, meaning that $\lambda$ is smaller than all other length scales of the system, the potential $\psi$ is large only at distances less than $\lambda$ from the charged walls. Far from the boundaries, the equations for the ionic concentrations and the potentials are considerably simplified, and the physical mechanisms at play in the electrostatic layer are taken into account via effective boundary conditions. A complete theory, including the flow induced by the motion of ions, was developed by Cox  in Ref.~\cite{cox1997electroviscous}. Since we will use this formalism as a starting point for our study, we recall it here but we refer to Ref.~\cite{cox1997electroviscous} for the derivation of this formalism.

First, due to the linearity of Stokes equation, we can separate the flow field into two contributions:
\begin{align}
\ve[v]=\ve[u] +\ve[u]^* 
\end{align}
where $\ve[u] $ is the flow induced by the motion of the lower wall only, \textit{i.e.} in the absence of ions, and $\ve[u]^*$ is the flow induced by the motion of ions, which satisfies Eq.~(\ref{4312}) but vanishes at the boundaries. 
 
Next, in the limit $\lambda\to0$, electroneutrality is achieved everywhere ($c_+=c_-=c$) except in the vicinity of the charged walls \cite{cox1997electroviscous}. At distances larger than $\lambda$ from the boundaries, the concentrations of cations and anions and the electric potential satisfy the bulk equations: 
\begin{align}
&\ve[u]  \cdot  {\boldsymbol{\nabla}} c = D_e \boldsymbol{\boldsymbol{\nabla}}^2 c,\label{AdvDiffc}\\
& \boldsymbol{\nabla}^2 \left(\frac{\psi \ q }{ k_\mathrm{B}T}+\beta \frac{c}{c_\infty} \right)=0,\label{EqPotentialCox}
\end{align}
these are equations (9.6), (9.7), (9.8) in Ref.~\cite{cox1997electroviscous}. 
This equation takes into account the fact that the flow $\ve[u]^*$ induced by the dynamics of ions is much lower than the imposed flow $\ve[u]$. Here,  $\beta$ is the relative difference of diffusivities between cations and anions, and $D_e$ is the effective ionic diffusivity:
\begin{align}
&\beta=\frac{D_+-D_-}{D_++D_-}, &D_e=\frac{2D_+D_-}{D_++D_-}.
\end{align}
Note that Eq. (\ref{EqPotentialCox}) expresses the fact that an electric field must spontaneously arise if there is an ionic concentration gradient, so that the \textcolor{black}{divergence of the} total charge current is zero \cite{prieve1984motion}. \textcolor{black}{It can also be obtained by subtracting   (\ref{AdvDiffPlus}) from (\ref{AdvDiffMinus}) assuming small deviations between $c=c_+=c_- $ and $c_\infty$.}
The effective boundary conditions  at the surfaces $\partial\Omega_\text{s}$, $ \text{s}\in\{\text{w},\text{p}\}$, are
\begin{align}
&\partial_n c= -\frac{2c_\infty \lambda^2 s_\perp^{\text{s}} }{D_e} \alpha_{\text{s}}, \label{BC_c}\\
& \partial_n \psi=\frac{ 2\lambda^2 s_\perp^{\text{s}} k_\mathrm{B}T  }{qD_e} \left[\Psi_{\text{s}} +4\beta \ln\cosh\left( \Psi_\text{s} /4\right)\right],  \label{BC_n}
\end{align}
see   equations (8.7), (8.9), (9.6) and (9.8) in Ref.~\cite{cox1997electroviscous}. 
Here, we recall that $\Psi_{\text{s}} =\psi_{\mathrm{s}}  q/k_\mathrm{B}T$ is the dimensionless potential at the surface $\partial\Omega_\text{s}$. The dimensionless parameter $\alpha_{\text{s}}$ depends only on the zeta potential of the surface and the difference of ionic diffusivities:
\begin{align}
\alpha_{\text{s}}=\beta   \Psi_{\text{s}} + 4 \ln \cosh\left( \Psi_{\text{s}}  /4 \right),
\end{align}
 and $s_\perp$ is defined near each solid boundary as
\begin{align}
s_\perp = \partial_n^2 (\ve[u] \cdot\ve[n])/2,\label{Def_s}
\end{align}
where $\partial_n=\ve[n]\cdot\boldsymbol{\nabla}$.  Eq. (\ref{BC_c}) means that $s_\perp$ can be considered as a source (or sink) term for the bulk ionic concentration, which is related to the perpendicular component of the velocity field in the EDLs.   Last, since electroneutrality holds in the bulk, the flow $\ve[u]^*$ induced by the motion of the ions satisfy the usual Stokes equations, 
\begin{align} 
&\boldsymbol{\nabla}\cdot\ve[u]^*=0, &
-{\boldsymbol{\nabla}} p^*+\eta{\boldsymbol{\nabla}}^2 \ve[u]^*=0.
\end{align}
However, in the electrostatic  Debye layer electroneutrality is broken, this induces an effective slip velocity $\ve[u]^*=\ve[u]_s^*$ at the boundaries (for the outer problem, \textit{i.e.}  at distances large compared to $\lambda$ from the interfaces, but small compared to  other length-scales), with 
\begin{align}
\ve[u]_s^*=\frac{2 c_\infty k_\mathrm{B}T \lambda^2}{\eta}\Big\{\Psi_\text{s} \boldsymbol{\nabla}_s \frac{\psi q } {k_\mathrm{B}T} -4 \ln \cosh  \frac{\Psi_\text{s}}{4} \boldsymbol{\nabla}_s  \frac{c}{c_\infty} 
   \Big\} 
   \label{slipv}.
\end{align}
This relation between the slip velocity and external gradients of concentration and electric potential can be found in Cox's  theory [see Eqs. (2.10) and (10.7) in Ref.~\cite{cox1997electroviscous}], and had been previously \textcolor{black}{obtained} in the context of diffusiophoresis and electrophoresis, see e.g. \cite{anderson1989colloid,prieve1984motion}.  
 Here, 
\begin{align}
\boldsymbol{\nabla}_s  =\boldsymbol{\nabla}  - \ve[n] \  \partial_n \label{DefNablaS}.
\end{align}

\section{Cox's equations in the lubrication regime $d\ll R$} 
\label{CoxLubrication}

Here, we present an analysis of Cox's equations, presented in Section \ref{CoxEquations}, in the lubrication regime $d\ll R$, with the condition $\lambda\ll d$  (in order to remain in the validity regime of Cox's equations). We define   $ l=\sqrt{R d}$ the characteristic length in the lateral direction, and we use rescaled coordinates
\begin{align}
&X=x/ l,  \hspace{1cm} Y=y/ l,  \hspace{1cm} Z=z/d,
\end{align}
so that the equation for the surface is
\begin{align}
Z=H(X,Y)\equiv 
1+\frac{1}{2}(X^2+Y^2) ,
\end{align}
where the dimensionless local distance  $H$ between the sphere and the surface is defined by the above equation. We introduce the small parameter $\varepsilon$ as
\begin{align}
\varepsilon=d/R=  (d/l)^2 \ll 1.
\end{align}

\subsection{Lubrication approximation for the flow induced by the motion of the wall}

Here we describe the flow $(u_x,u_y,u_z)$, associated with the pressure field $p$, generated by the motion of the wall at velocity $V$, in the absence of ions.  This is a standard hydrodynamic problem \cite{oneill1967slow,jeffrey1981slow}. In lubrication theory we have the scaling
\begin{align}
&u_x=V U_X(X,Y,Z),&
u_y=V U_Y(X,Y,Z), \label{Equxuy}\\
&u_z= V \varepsilon^{1/2} U_Z(X,Y,Z),
&p=\frac{\eta V l }{d^2}P(X,Y,Z). \label{Equz}
\end{align}
Inserting this ansatz into Stokes equations, we obtain   
\begin{align}
&-\boldsymbol{\nabla}_{\parallel}P+\partial_Z^2\ve[U]_\parallel=0, &- \partial_Z P =0,\label{7432} \\ 
&\boldsymbol{\nabla}_{\parallel}\cdot\ve[U]_\parallel+\partial_ZU_Z=0, \label{COnt9432} 
\end{align}
where 
$\ve[U]_\parallel=U_X\hat{\ve[e]}_x+U_Y\hat{\ve[e]}_y$ and $\boldsymbol{\nabla}_{\parallel}=\partial_X \hat{\ve[e]}_x+\partial_Y \hat{\ve[e]}_x$. Note that $\boldsymbol{\nabla}_\parallel$ should not be confounded with  the projection $\boldsymbol{\nabla}_s$ [see Eq.~(\ref{DefNablaS})] of $\boldsymbol{\nabla}$ along the boundaries, since these ones are not exactly horizontal. Noting that $P$ does not depend on $Z$, the solution to Eq. (\ref{7432}) reads 
\begin{align}
&\ve[U]_\parallel=\frac{H-Z}{H}\hat{\ve[e]}_x+(\boldsymbol{\nabla}_\parallel P)\frac{ Z(Z-H)}{2}\label{Uparallel}.
\end{align}
Inserting this expression into the continuity equation (\ref{COnt9432}), using $U_Z(0)=0$ and integrating over $Z$ leads to
\begin{align}
U_Z=   -  \frac{Z^2\partial_XH}{2H^2}+ \boldsymbol{\nabla}_\parallel\cdot\left[(\boldsymbol{\nabla}_\parallel P)\left(-\frac{Z^3}{6}+\frac{Z^2H}{4}\right)\right].
\end{align}
The condition   $U_Z(H)=0$ at the second boundary leads to \textcolor{black}{the Reynolds} equation for the pressure
\begin{align}
 \frac{\partial_XH}{2}= \boldsymbol{\nabla}_\parallel\cdot\left[(\boldsymbol{\nabla}_\parallel P) \frac{H^3}{12}\right].\label{EqPressureHydro}
\end{align}

The only physically acceptable solution is \cite{oneill1967slow}
 \begin{align}
&P=-6 X/(5H^2)  .\label{Pressure}
\end{align}

Now,  the  source terms   $s_\perp$, defined in Eq. (\ref{Def_s}), can be \textcolor{black}{calculated}. The normal pointing towards the liquid at the sphere's surface is, at next-to-leading order in $\sqrt{\varepsilon}$,
\begin{align}
\ve[n]
\simeq -\hat{\ve[e]}_z+\sqrt{\varepsilon}\boldsymbol{\nabla}_{\parallel}H.
\end{align}
Note that the term proportional to $\boldsymbol{\nabla}_{\parallel}H$ indicates that the normal is not exactly oriented along the vertical direction. Using (\ref{Equxuy}), and (\ref{Equz}), we obtain
\begin{align}
\ve[n]\cdot \ve[u]=V \sqrt{\varepsilon} \ (-U_Z+ \boldsymbol{\nabla}_\parallel H\cdot\ve[U]_\parallel ).
\end{align}
\textcolor{black}{Inserting} this expression into Eq. (\ref{Def_s}), we obtain 
\begin{align}
s_\perp^{\text{p}} =\frac{V \sqrt{\varepsilon} }{d^2}S_\perp^{\text{p}} (X,Y),\label{Scaling_f}
\end{align}
where $S_\perp^{\text{p}} $ can be identified in terms of the rescaled flow at the sphere's surface:
\begin{align}
S_\perp ^{\text{p}}(X,Y)=\lim_{Z\to H}\frac{-U_Z+ \boldsymbol{\nabla}_\parallel H\cdot\ve[U]_\parallel }{(H-Z)^2}. \label{FperpS}
\end{align}
For the wall, we follow the same reasoning with $\ve[n]=\hat{\ve[e]}_z$, leading to 
\begin{align}
S_\perp ^{\text{w}}(X,Y)=\lim_{Z\to 0}U_Z /Z^2. \label{FperpW}
\end{align}
In the three-dimensional case, the values $S_\perp ^s$ are
\begin{align}
&S_\perp ^{\text{w}}=-\frac{2 X \left(X^2+Y^2-4\right)}{5 H^3 }, \label{Sperp3DW}\\
&S_\perp ^{\text{p}}=\frac{ X \left(X^2+Y^2+26\right)}{10 H^3}.\label{Sperp3DS}
\end{align}
 
\subsection{Lubrication approximation for the ionic concentration $c$}

Let us introduce the Péclet number associated to the length scale $ l$, defined as
\begin{align}
\text{Pe}   
=\frac{V \sqrt{d R}}{  D_e } \textcolor{black}{=\frac{V l }{  D_e } },
\end{align}
so that the advection diffusion Eq.~(\ref{AdvDiffc}) for $c$ becomes
\begin{align}
\text{Pe}\ \varepsilon  (U_X\partial_X c +U_Y \partial_Yc+U_Z \partial_Z c)= \partial_Z^2c +\varepsilon \boldsymbol{\nabla}_{\parallel}^2 c  . \label{64351}
\end{align}
\textcolor{black}{In the following, } we keep the parameter Pe constant when taking the limit $d/R\to0$. 
For the boundary conditions at the sphere's surface, we note that, for any field $\phi$, 
\begin{align}
\ve[n]\cdot \boldsymbol{\nabla} \phi = \frac{1}{d}[-\partial_Z\phi+  \varepsilon (\boldsymbol{\nabla}_\parallel H)\cdot (\boldsymbol{\nabla}_\parallel \phi) ]. \label{GenFormNablan}
\end{align}
Using the above expression and Eq.~(\ref{Scaling_f}),    the boundary condition Eq.~(\ref{BC_c}) becomes
\begin{align}
  & [-\partial_Zc+  \varepsilon (\boldsymbol{\nabla}_\parallel H)\cdot \boldsymbol{\nabla}_\parallel c) ]_{Z=H}    =-  \frac{2c_\infty \lambda^2 V   }{\sqrt{d R} D_e}   S_\perp ^{\text{p}}       \alpha_{\text{p}}, \label{64352} \\
 & (\partial_Zc)_{Z=0}    = -2 \frac{c_\infty \lambda^2 V  }{d^{1/2}\sqrt{R} D_e}  S_\perp ^{\text{w}}  \alpha_{\text{w}}.\label{64353}
 \end{align}

Equations (\ref{64351}), (\ref{64352}), (\ref{64353}) suggest   looking for solutions under the form
\begin{align}
c =c_\infty+\frac{c_\infty \lambda^2   }{d^{2}  }    (N+\varepsilon N_1+...),\label{Expc}
\end{align}
so that, at leading order in $\varepsilon$, 
\begin{align}
&\partial_Z^2N=0, &(\partial_Z N)_{Z=0}=(\partial_Z N)_{Z=H}=0,
\end{align}
which means that $N $ depends only on $X,Y$. At next order we have
\begin{align}
&\text{Pe} \ \ve[U]_\parallel  \cdot\boldsymbol{\nabla}_\parallel  N    = \partial_Z^2N_1 + \boldsymbol{\nabla}_{\parallel}^2 N , \label{0421}\\
&(-\partial_Z N_1+\boldsymbol{\nabla}_{\parallel} H \cdot \boldsymbol{\nabla}_{\parallel} N )_{Z=H}=-2\alpha_{\text{p}} \text{Pe} \ S_\perp ^{\text{p}}, \label{95842}\\
&(\partial_Z N_1)_{Z=0}=-2\alpha_{\text{w}} \text{Pe} \ S_\perp ^{\text{w}}.\label{95843}
\end{align}
Integrating Eq.~(\ref{0421}) over $Z\in[0,H]$ and using the boundary conditions Eqs.~(\ref{95842}) and (\ref{95843}), we obtain  
  \begin{align}
   \text{Pe} \overline{\ve[U]}_\parallel \cdot\boldsymbol{\nabla}_{\parallel} N  = 
   &\boldsymbol{\nabla}_{\parallel}\cdot (H\boldsymbol{\nabla}_{\parallel}  N )
+2\text{Pe}  (\alpha_{\text{w}}S_\perp ^{\text{w}}+\alpha_{\text{p}} S_\perp ^{\text{p}}). \label{EqN}
\end{align}
where $\overline{\ve[U]}_\parallel=\int_0^H dZ   {\ve[U]}_\parallel$. $N $ satisfies an effective stationary advection-diffusion equation, with an effective velocity field $\overline{\ve[U]}_\parallel$ that is incompressible due to Eq.~(\ref{EqPressureHydro}), and an external source term. 
The effective flow $ \overline{\ve[U]}_\parallel$ can be calculated from  Eq.~(\ref{Uparallel}), so that
\begin{align}
 \overline{\ve[U]}_\parallel=
 \frac{H}{2}   \hat{\ve[e]}_x & -\frac{H^3}{12}  \boldsymbol{\nabla}_{\parallel} P . 
 \end{align}
Using (\ref{Pressure}), we obtain 
\begin{align}
&\overline{U}_X = \frac{3}{5} + \frac{X^2+3Y^2}{10},   \ \ \
\overline{U}_Y =-\frac{ X Y}{5}.  \label{UBar3D}
\end{align}

\subsection{Lubrication approximation for the electric potential}

Next, we seek an effective equation for the electric potential. Eq.~(\ref{EqPotentialCox}) suggests that it is simpler to work with the reduced potential $ \mathcal{A}$ defined as
\begin{align}
 \mathcal{A} =\psi+ \beta  \frac{k_\mathrm{B}T }{q c_\infty} c,  \label{85431}
\end{align}
which satisfies $\boldsymbol{\nabla}^2 \mathcal{A}=0 $. 
The boundary conditions for $ \mathcal{A}$ are obtained by combining Eqs.~(\ref{BC_c}) and (\ref{BC_n}) :
\begin{align}
& \partial_n  \mathcal{A}=2\frac{\lambda^2 k_\mathrm{B}T s_\perp^{\text{s}}}{D_e q} (1-\beta^2)\Psi_\text{s}. 
\end{align}
Using Eqs. (\ref{GenFormNablan}) and (\ref{Scaling_f}), we can write
\begin{align}
 & (-\partial_Z\mathcal{A}+  \varepsilon \boldsymbol{\nabla}_\parallel H \cdot \boldsymbol{\nabla}_\parallel \mathcal{A})_{H}    =  \frac{2k_\text{B}T  \lambda^2 \text{Pe}   S_\perp ^{\text{p}} (1-\beta^2)\Psi_\text{p}   }{q d^2 }\varepsilon        \\
 & (\partial_Z\mathcal{A})_{Z=0}    =  \frac{2k_\text{B}T  \lambda^2 \text{Pe}   S_\perp ^{\text{w}} (1-\beta^2)\Psi_\text{w}   }{q d^2 }\varepsilon  .
 \end{align}
We look for solutions scaling as 
\begin{align}
 \mathcal{A}= \frac{k_\mathrm{B}T \lambda^2   }{q d^{2}  }  (A+\varepsilon A_1+...)\label{85432}
\end{align}
The equations for $A$ and $A_1$ are exactly the same as those for $N $ and $N_1$ in the absence of advective term, if one replaces the factors $-\alpha_\text{s}$ by $(1-\beta^2)\Psi_\text{s}$. We thus directly write 
\begin{align}
 \boldsymbol{\nabla}_{\parallel}\cdot(H\boldsymbol{\nabla}_{\parallel}  A )  =2(1-\beta^2) \text{Pe}  \left(\Psi_\text{p}S_\perp ^{\text{p}}+\Psi_{\text{w}}S_\perp ^{\text{w}}\right) \label{EqA}.
\end{align}

\subsection{Lubrication approximation for the flow induced by the motion of ions}

We now extend the lubrication approximation to the flow $\ve[u]^*$ induced by the motion of ions. First, we note that surface nabla operator $\boldsymbol{\nabla}_s$, given by Eq.~(\ref{DefNablaS}), when applied to $c$ given by Eq.~(\ref{Expc}) at the sphere's surface, reads 
\begin{align}
\boldsymbol{\nabla}_s c=\frac{c_\infty \lambda^2     }{d^{2}   l}  \left[\boldsymbol{\nabla}_{\parallel}N +(\boldsymbol{\nabla}_{\parallel}N \cdot \boldsymbol{\nabla}_{\parallel}H)\sqrt{ \varepsilon} \hat{\ve[e]}_z +\mathcal{O}(\varepsilon)\right],
\end{align}
where we have used the fact that $N $ depends only on $X,Y$. This expression is valid at the surface of the particle, while at the wall the same expression holds without the term $\boldsymbol{\nabla}_\parallel H$. Note that, in the above expression, the vertical component comes from the fact that the tangent to the surface is not exactly in the horizontal plane.  
Using this result and a similar expression for the potential, we see that the slip velocity in Eq.~(\ref{slipv}), at the sphere's surface, reads
\begin{align}
\ve[u]_p^*&=v^* \Big\{-\alpha_{\mathrm{p}} \boldsymbol{\nabla}_{\parallel} N  +\Psi_{\mathrm{p}} \boldsymbol{\nabla}_{\parallel}  A \nonumber\\
&+\sqrt{\varepsilon}(\boldsymbol{\nabla}_\parallel H)\cdot \left(  - \alpha_{\mathrm{p}} \boldsymbol{\nabla}_{\parallel} N  +\Psi_{\mathrm{p}} \boldsymbol{\nabla}_{\parallel}  A    \right)\hat{\ve[e]}_z \Big\},\label{SlipS}
\end{align}
with the characteristic induced velocity
\begin{align}
v^*=\frac{2 c_\infty k_\mathrm{B}T \lambda^4}{\eta d^{2}  l}=\frac{\epsilon (k_\mathrm{B}T)^2 \lambda^2 }{\eta q^2 d^{2}  l }.\label{CharSlip}
\end{align}
Similarly, at the wall, the slip velocity is
\begin{align}
\ve[u]_\text{w}^*=v^*& \left\{-\alpha_{\mathrm{w}} \boldsymbol{\nabla}_{\parallel} N  +\Psi_{\mathrm{w}} \boldsymbol{\nabla}_{\parallel}  A  \right\},\label{SlipW}
\end{align}

Choosing $v^*$ as the velocity scale for the flow induced by the ions, we have the following structure for this flow in the lubrication regime:
\begin{align}
&u_x^*=v^* U_X^*, &u_y^*=v^* U_Y^*\\
&u_z^*=v^*\sqrt{\varepsilon}U_Z^*, &p^*= \frac{\eta v^* l}{d^2}P^*,
\end{align}
\textcolor{black}{where $p^*$ is the pressure associated to $\ve[u]^*$. }
The equations for the rescaled induced flow are
\begin{align}
&-\boldsymbol{\nabla}_{\parallel}P^*+\partial_Z^2\ve[U]_\parallel^*=0, &\partial_ZP^*=0 \label{09421}, \\
&\boldsymbol{\nabla}_{\parallel}\cdot\ve[U]_\parallel^*+\partial_ZU_Z^*=0.  \label{COntStar}
\end{align}
The associated boundary conditions are obtained from Eqs.~(\ref{SlipS}) and (\ref{SlipW}) and read
\begin{align}
&( \ve[U]_\parallel^*)_{Z=0}=-\alpha_{\text{w}}\boldsymbol{\nabla}_{\parallel} N  +\Psi_{\text{w}} \boldsymbol{\nabla}_{\parallel}  A  ,\\
&( \ve[U]_\parallel^*)_{Z=H}=-\alpha_{\text{p}}\boldsymbol{\nabla}_{\parallel} N  +\Psi_\text{p} \boldsymbol{\nabla}_{\parallel}  A , \\
&(U_Z^*)_{Z=0}=0,\\
&(U_Z^*)_{Z=H}={\boldsymbol{\nabla}_\parallel}H\cdot \left(  -\alpha_{\text{p}} \boldsymbol{\nabla}_{\parallel} N  +\Psi_{\text{p}} \boldsymbol{\nabla}_{\parallel}  A  \right). \label{VzH}
\end{align}

Noting that $P^*$ does not depend on $Z$, integrating Eq.~(\ref{09421}) over $Z$ and using the above boundary conditions leads to
\begin{align}
\ve[U]_\parallel^*=&\frac{(\boldsymbol{\nabla}_{\parallel} P^*)Z(Z-H)}{2}-(\alpha_{\text{w}} \boldsymbol{\nabla}_{\parallel} N  -\Psi_{\text{w}} \boldsymbol{\nabla}_{\parallel} A )\frac{H-Z}{H} \nonumber\\
&-  (\alpha_{\text{p}} \boldsymbol{\nabla}_{\parallel} N  -\Psi_\text{p} \boldsymbol{\nabla}_{\parallel}  A )\frac{Z}{H}.
\end{align}
We insert this expression into the continuity equation Eq.~(\ref{COntStar}) and integrate the resulting equation over $Z\in [0,H]$; with the use of Eq.~(\ref{VzH}) this leads to
\begin{align}
\frac{ \boldsymbol{\nabla}_\parallel\cdot (H^3\boldsymbol{\nabla}_{\parallel} P^*)}{12} = &
      -\frac{ 1}{2}  (\alpha_{\text{w}}+\alpha_{\text{s}})\boldsymbol{\nabla}_{\parallel}\cdot(H\boldsymbol{\nabla}_{\parallel} N ) \nonumber\\
&      +\frac{\Psi_{\text{w}}+\Psi_\text{p}}{2} \boldsymbol{\nabla}_{\parallel}\cdot(H\boldsymbol{\nabla}_{\parallel}A ). \label{EqPstar}
   \end{align}   
 This equation, with the condition $P^*\to0$ at infinity, defines the pressure $P^*$ associated to the induced flow. 

\subsection{Expression of the electroviscous lift force}

The lift force can be evaluated at $Z=0$ (above the electrostatic Debye layer). 
\textcolor{black}{We note that, at leading order, the integral of the pressure (\ref{Pressure}), induced by the motion of the wall, vanishes. This recovers the well-known result that there is no hydrodynamic lift at zero Reynolds number. }
We find that, at \textcolor{black}{next-to-}leading order, the lift force comes from the pressure associated to the flow $\ve[u]^*$ induced by the motion of ions:
\begin{align}
&F_\text{lift} = \int d\ve[r]\  p^*=  \frac{\eta v^*  l^3}{d^2}\int d\ve[R]_\parallel P^*, 
\end{align}
where $d\ve[R]_\parallel =dX dY$.  

We define a dimensionless force
\begin{align}
\Phi(\text{Pe})= \int d\ve[R]_\parallel  P^*(\ve[R]_\parallel ). \label{DefFTilde}
\end{align}
With this definition, the lift force reads
\begin{align}
&F_\text{lift} = \epsilon \left(\frac{k_\mathrm{B}T \lambda}{qd^2} \right)^2             l^2   \  \Phi(\text{Pe})    . \label{MainContr}
\end{align}

We also define the Maxwell (M) component of the force, whose leading order component reads [see Eqs.~(\ref{85431}), (\ref{85432}) and (\ref{StressGen})]:
\begin{align}
F_\text{lift}^{\text{M}} =\epsilon\left(\frac{k_\mathrm{B}T\lambda^2}{q d^2}\right)^2 \frac{   l^{3}}{2l^3    }   \int  d\ve[R]_\parallel   [\boldsymbol{\nabla}_\parallel (A -\beta N )]^2.
\end{align}
It is instructive to compare the magnitude of this Maxwell contribution to the main contribution (\ref{MainContr})
\begin{align}
\frac{F_\text{lift}^{\text{M}}}{F_\text{lift} }\sim \frac{\lambda^2    }{ Rd   } \ll 1,
\end{align}
meaning that, when $\lambda\ll \sqrt{Rd}$,   the viscous contribution to the  lift force dominates over the contribution of Maxwell's stress. 

Let us simplify the expression for the force. First, an integration by part in Eq.~(\ref{DefFTilde}) leads to
\begin{align}
\Phi =-\frac{1}{ 2}\int d\ve[R]_\parallel\ \ve[R]_\parallel\cdot\boldsymbol{\nabla}_\parallel P^*(\ve[R]_\parallel ),
\end{align}
where we have assumed that $P^*$ decays fast enough for large $\vert\ve[R]_\parallel\vert$ (as is the case in practice). Note that we have used $\boldsymbol{\nabla}_\parallel \cdot \ve[R]_\parallel=2$.  
For a parabolic surface, we have $\ve[R]_\parallel=\boldsymbol{\nabla}_\parallel H$, so that we can write
\begin{align}
\Phi  =-\frac{1}{2}\int d\ve[R]_\parallel \frac{\boldsymbol{\nabla}_\parallel H }{H^3}\cdot (H^3\boldsymbol{\nabla}_\parallel P^* ). 
\end{align}
Applying again the divergence formula leads to
\begin{align}
\Phi =-\frac{1}{4}\int d\ve[R]_\parallel \frac{\boldsymbol{\nabla}_\parallel\cdot(H^3\boldsymbol{\nabla}_\parallel P^* )}{H^2}  . \label{9594}
\end{align}
Using Eq. (\ref{EqPstar})   for the rescaled pressure, we obtain
\begin{align}
\Phi  =\frac{3}{2}\int &\frac{d\ve[R]_\parallel }{H^2}  \Big[  (\alpha_{\text{w}}+\alpha_{\text{p}})\boldsymbol{\nabla}_{\parallel}\cdot(H\boldsymbol{\nabla}_{\parallel} N ) \nonumber\\
&      - (\Psi_{\text{w}}+\Psi_{\text{s}}) \boldsymbol{\nabla}_{\parallel}\cdot(H\boldsymbol{\nabla}_{\parallel}A ) \Big] .\label{89R5442}
\end{align}
The term involving $A$ can then be simplified by using Eq.~(\ref{EqA}) and one can easily check that 
\begin{align}
& \int d\ve[R]_\parallel \frac{S_\perp ^{\text{w}}}{H^2} =\int d\ve[R]_\parallel \frac{S_\perp ^{\text{p}}}{H^2}  =0,\label{Trick}
\end{align}
so that the contribution of the term $A$ to the force is exactly zero. 
Ignoring the contribution of the term $A$ in Eq.~(\ref{89R5442}) and applying once again the divergence formula, we obtain
\begin{align}
\Phi =&3(\alpha_{\text{w}}+\alpha_{\text{p}}) \int d\ve[R]_\parallel \frac{(\boldsymbol{\nabla}_{\parallel}H)\cdot(\boldsymbol{\nabla}_{\parallel} N )}{H^2}      . \label{1034} 
\end{align}
We may also write an alternative form of this rescaled force by using Eqs. (\ref{89R5442}), (\ref{EqN}) and  (\ref{Trick}): 
\begin{align}
\Phi =& \frac{3 \text{Pe}\ (\alpha_{\text{w}}+\alpha_{\text{p}})}{2} \int  \frac{d\ve[R]_\parallel}{H^2} \   \overline{\ve[U]}_\parallel\cdot \boldsymbol{\nabla}_{\parallel} N  . \label{EQ01334}
\end{align}
In summary, in the regime $\lambda\ll d\ll R$, the lift force is given by Eqs.~(\ref{MainContr}) and (\ref{1034}), where $N$ is the solution of Eq.~(\ref{EqN}).
  
\section{Explicit solutions}
\label{SecExplicitSol3D}

\subsection{Structure of the solution}

\begin{figure*}[ht!]
    \centering
    \includegraphics[width=16cm]{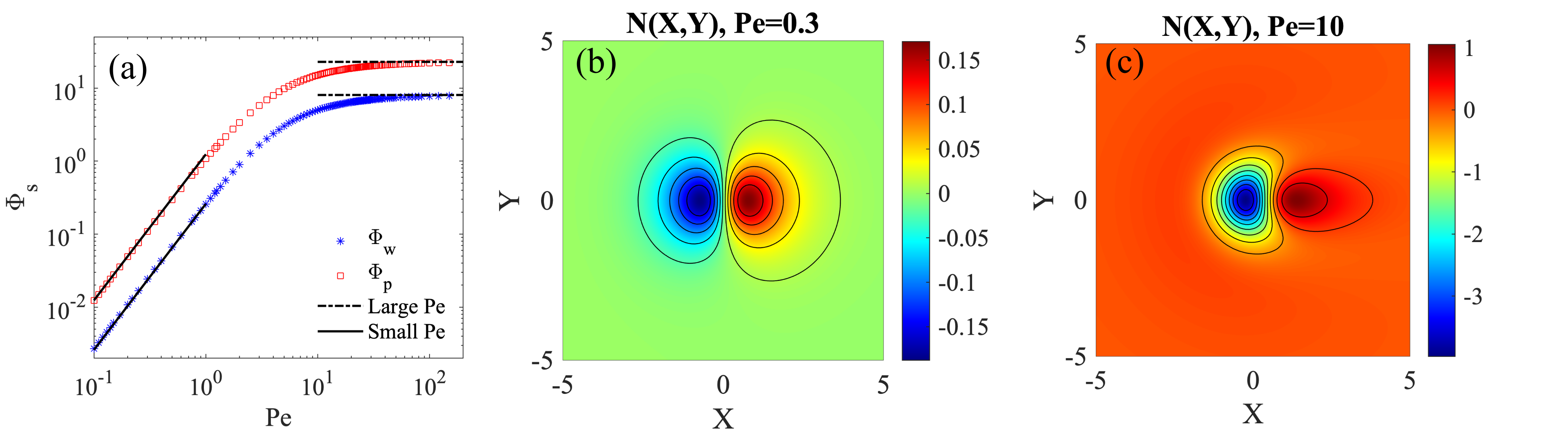}
    \caption{(a) $\Phi_\text{p}$ and $\Phi_\text{w}$ as a function of the Péclet number. Symbols are the results of the numerical integration of Eqs.~(\ref{DefPHIp}) and (\ref{EqNp}).  Full lines represent the retained values for small Péclet, given in Eq.~(\ref{OurValuesAtZero}). Dashed lines represent the analytical results  (\ref{PHIsLargePe}) and (\ref{PHIwLargePe}) for large Péclet numbers.  (b) Dimensionless concentration $N=\alpha_p N_p +\alpha_w N_w$ for $\text{Pe}=0.3$, where the $\alpha_i$ are evaluated using the experimental values of the surface potentials and the diffusivities of Na$^+$ and Cl$^-$, obtained from the numerical integration  of Eqs.~(\ref{DefPHIp}) and (\ref{EqNp}) over a disk of radius $8$ (in units of $l$). (c) Same as (b) except for $\text{Pe}=10$.}   
    \label{fig:PHIp}
\end{figure*}

Looking at Eqs.~(\ref{1034}) and (\ref{EqN}), it is clear that the dimensionless force can be put under the form
\begin{align}
\Phi =(\alpha_{\mathrm{p}}+\alpha_{\mathrm{w}})(\Phi_\text{w} \alpha_{\mathrm{w}}+\Phi_\text{p} \alpha_{\mathrm{p}}), \label{FoncForm}
\end{align}
where 
\begin{align}
&\Phi_\mathrm{s} = 3 \int d\ve[R]_\parallel \frac{\boldsymbol{\nabla}_{\parallel}H \cdot \boldsymbol{\nabla}_{\parallel} N_\mathrm{s} }{H^2},  & \text{s}\in\{\text{p},\text{w}\}, \label{DefPHIp}
\end{align}
and $N_\mathrm{s}$ is the solution of Eq.~(\ref{EqN}) when only the source term corresponding to the surface $\mathrm{s}$ is present, \textit{i.e.} $N_\mathrm{s}$ satisfies
  \begin{align}
   \text{Pe}\ \overline{\ve[U]}_\parallel\cdot\boldsymbol{\nabla}_\parallel  N_\mathrm{s}=
   \boldsymbol{\nabla}_{\parallel}\cdot (H\boldsymbol{\nabla}_{\parallel}  N_\mathrm{s}  )
+2\text{Pe}\  S_\perp ^{\text{s}}. \label{EqNp}
\end{align}
Both functions $\Phi_\text{w}$ and $\Phi_\text{p}$ depend only on the Péclet number and not on the electrostatic properties of the surfaces. 
These functions can be obtained numerically using a solver for partial differential equations. They are represented in Fig.~\ref{fig:PHIp}. 

\subsection{Limit of small Péclet number}
\subsubsection{Case of equal zeta potentials at the particle and the wall}

Analytical results can be found in the case that the wall and the surface have the same zeta potential, so that $\alpha_{\mathrm{w}}=\alpha_{\mathrm{p}}\equiv\alpha$). 
In this case, at small $\text{Pe}$, if we expand $N$ in powers of the Péclet number as $N\simeq  N_0+\text{Pe}\ N_1+...$, we readily obtain that $N_0$ is zero, while $N_1$ satisfies 
\begin{align}
&H  \boldsymbol{\nabla}_{\parallel}^2 N_1   +\boldsymbol{\nabla}_{\parallel} H\cdot \boldsymbol{\nabla}_{\parallel} N_1 =-2\alpha(S_\perp^{\text{w}} +S_\perp^{\text{s}}).\end{align}
The solution to this equation was given in Ref. \cite{tabatabaei2006electroviscousSphere}
\begin{align}
N_1=\frac{6 \alpha X}{5 H^2 }.
\end{align}
Inserting the value of $N\simeq \text{Pe} N_1$ into Eq.~(\ref{EQ01334}), we find
\begin{align}
\Phi=\frac{ 24 \alpha^2 \pi}{25} \text{Pe}^2.\label{FSmallPécletEqPotential}
\end{align}
This result is exactly that obtained in Refs.~\cite{tabatabaei2006electroviscousSphere} [see equation (6.27) in this reference] and \cite{schnitzer2016streaming} for small Péclet numbers in the same conditions. 

\subsubsection{Case of distinct zeta potentials}
\label{SecSmallV3DGen}
In the case that the wall and the sphere have different electrostatic properties, we could not find analytical expressions for the force in the limit of small Péclet number. However, it is clear that one can still write $N=\text{Pe}N_1$ for small Péclet, where $N_1$ satisfies
 \begin{align}
   0 = \boldsymbol{\nabla}_{\parallel}\cdot (H\boldsymbol{\nabla}_{\parallel}  N_1 )
+2   (\alpha_{\text{w}}S_\perp ^{\text{w}}+\alpha_{\text{p}} S_\perp ^{\text{p}}). \label{EqN1}
\end{align}
Numerically solving for $N_1$ leads to
\begin{align}
&\Phi_\text{w} \underset{\text{Pe}\to0}{\simeq} 0.26 \text{Pe}^2, &\Phi_\text{p}\underset{\text{Pe}\to0}{\simeq}1.24 \ \text{Pe}^2. \label{OurValuesAtZero}
\end{align}
Note that, with these values, one recovers  Eq.~(\ref{FSmallPécletEqPotential}) when $\alpha_\text{p}=\alpha_\text{w}$. Interestingly, the result of Ref.~\cite{tabatabaei2006electroviscousSphere}, as given by their equation (6.27), can be written as 
\begin{align}
\Phi=\frac{6\pi(\alpha_{\mathrm{p}}+\alpha_{\mathrm{w}})}{5}\left[\frac{\alpha_{\mathrm{w}}+\alpha_{\mathrm{p}}}{5}+1.667(\alpha_{\mathrm{p}}-\alpha_{\mathrm{w}}) \right]\text{Pe}^2.
\end{align}
This corresponds to our functional form Eq.~(\ref{FoncForm}) for small Péclet, with 
\begin{align}
&\Phi_\text{w} \underset{\text{Pe}\to0}{\simeq}  -5.53 \ \text{Pe}^2, 
&\Phi_\text{p} \underset{\text{Pe}\to0}{\simeq}  7.03\ \text{Pe}^2,
\end{align}
which  differ from our results in Eq.~(\ref{OurValuesAtZero}). This discrepancy may come from the fact that one of the steps  in Ref.~\cite{tabatabaei2006electroviscousSphere} consists in finding the solution to Eq. (\ref{EqN1}), which is done by writing $N_1$ in terms of an infinite series [see Eqs. (4.33)-(4.35) in Ref. \cite{tabatabaei2006electroviscousSphere}]. However, this series converges to a function that is divergent at the origin and is thus not an acceptable solution for $N_1$.

\subsection{Large Péclet numbers}
In the limit of large Péclet numbers, $N$ becomes independent on the Péclet number, and the diffusive term can be neglected in Eq. (\ref{EqN}) which becomes 
  \begin{align}
   \overline{\ve[U]}_\parallel \cdot\boldsymbol{\nabla}_{\parallel} N  = 
2   (\alpha_{\text{w}}S_\perp ^{\text{w}}+\alpha_{\text{p}} S_\perp ^{\text{p}}).  
\end{align}
Using the expressions of the source terms and the effective flow [see Eqs (\ref{Sperp3DW}), (\ref{Sperp3DS}) and (\ref{UBar3D})] and using  polar coordinates with $X=\rho\cos\theta$ and $Y=\rho\sin\theta$, we obtain 
\begin{align}
\frac{6 + \rho^2}{10} & \cos\theta \ \partial_\rho N -\frac{6 + 3\rho^2}{10\ \rho} \sin\theta\ \partial_\theta N=\nonumber\\
2\cos\theta&\left[\alpha_\text{w}\frac{16 \rho (4-\rho^2) }{5 (\rho^2 +2 )^3}+ \alpha_\text{p} \frac{4 \rho \left(\rho^2+26\right)  }{5 \left(\rho^2+2\right)^3}\right].
\end{align}
We remark that we can find a solution $N(\rho)$ that depends only on the radial coordinate $ \rho$, so that
\begin{align}
    \partial_\rho N  =
16 \ \frac{4 \alpha_\text{w}  \rho (4-\rho^2) + \alpha_\text{p}   \rho \left(\rho^2+26\right)  }{(6 + \rho^2)\left(\rho^2+2\right)^3}   .
\end{align}
Using this expression to calculate Eq.~(\ref{1034}), we find that the dimensionless lift force for large Péclet number reads
\begin{align}
&\Phi  \simeq
\pi (\alpha_\text{w}+\alpha_\text{p})\left[  (64-45 \ln 3)  \frac{\alpha_\text{p}}{2  }  
+\alpha_\text{w}  (52-45 \ln3)\right] .\label{PHILargePe}
\end{align}
 Equivalently, we can write the limiting values of $\Phi_\text{p},\Phi_\text{w}$ for large Péclet as
\begin{align}
&\Phi_\text{p}\underset{\text{Pe}\to\infty}{\simeq} \pi  (64-45 \ln 3)  \frac{1}{2  }\simeq 22.87,\label{PHIsLargePe}
\\
&\Phi_\text{w}\underset{\text{Pe}\to\infty}{\simeq}\pi  (52-45 \ln3) \simeq 8.05 .\label{PHIwLargePe}
\end{align}

\begin{figure}[ht!]
    \centering
    \includegraphics[width=8cm]{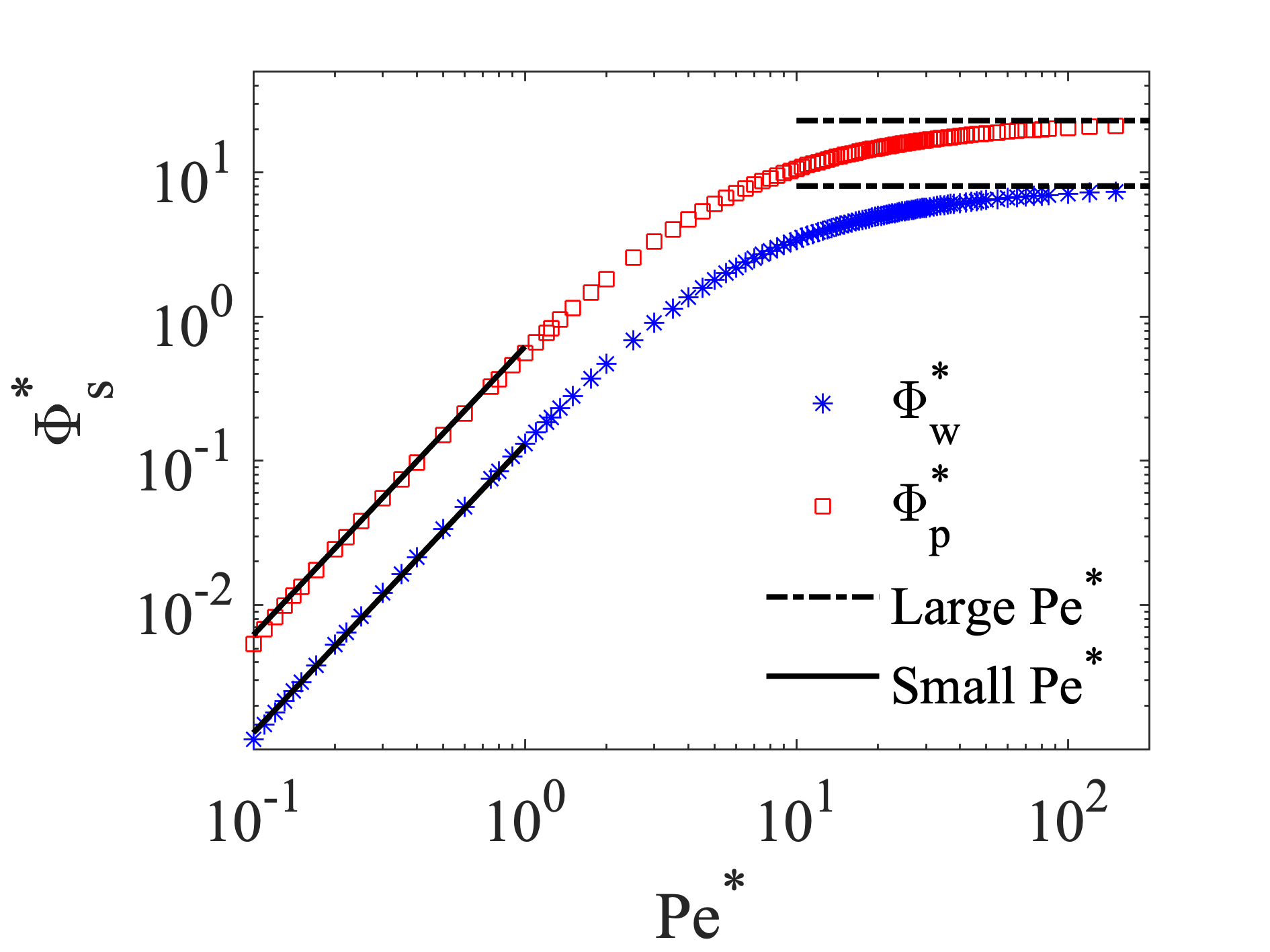}
    \caption{$\Phi_\text{p}^*$ and $\Phi_\text{w}^*$ as a function of the characteristic Péclet number $\text{Pe}_*$  in the experimentally relevant case of sinusoidal variation of the velocity. Symbols are obtained by using Eq.~(\ref{PHIStar}) and the data for $\Phi_\text{s}$ and $\Phi_\text{w}$ shown in Fig.~\ref{fig:PHIp}. Full lines represent the retained values for small Péclet, given in Eq.~(\ref{OurValuesAtZero}), with $\Phi_\text{s}^*\simeq\Phi_\text{s}/2$. Dashed lines represent the analytical results Eqs.~(\ref{PHIsLargePe}) and (\ref{PHIwLargePe}) for large Péclet numbers, where $\Phi_\text{s}^*\simeq\Phi_\text{s}$. }
    \label{fig:PHIpStar}
\end{figure}

\subsection{Temporally averaged force}
In the experiments the velocity is not constant but is applied as a sinusoidal function:
\begin{align}
V(t)=V_0 \sin(\omega t)
\end{align}
so that the (instantaneous) Péclet number becomes time dependent. We therefore define
\begin{align}
&\text{Pe}_*=V_0 \sqrt{R d}/D_e, &\text{Pe}(t)=\text{Pe}_*\sin(\omega t).
\end{align}
If $\omega$ is sufficiently low, the instantaneous force predicted by our theory reads
\begin{align}
&F_\text{lift}= F_0  \ \Phi(\text{Pe}_*\sin(\omega t)) .
\end{align}
Here we have assumed that $\omega$ is sufficiently low so that one can consider that the force depends only on the instantaneous velocity. The validity of this quasi-static approximation is discussed in Section \ref{SectionDyn}.
The average force can be written as 
\begin{align}
&\langle F_\text{lift}\rangle=   F_0 \  \Phi^*(\text{Pe}_*),
\end{align}
where the scaling function $\Phi^*$ reads
\begin{align}
 \Phi^*(\text{Pe}_*)=\frac{1}{2\pi}\int_0^{2\pi} dt   \ \Phi(\text{Pe}_*\sin t).
\end{align}
 In practice we may evaluate this integral  on the interval $[0,\pi/2]$ only, by using   $u=\text{Pe}_*\sin(t)$, one obtains
\begin{align}
& \Phi^*(\text{Pe}_*)
= \frac{2}{\pi }  \int_0^{\text{Pe}_*} \frac{du}{\text{Pe}_* \sqrt{1-\left(\frac{u}{\text{Pe}_*}\right)^2}} \Phi(u).
\end{align}
We may also define $\Phi_\text{s}^*$ such that 
\begin{align}
\Phi^*=(\alpha_{\mathrm{p}}+\alpha_{\mathrm{w}})(\Phi_\text{w}^* \alpha_{\mathrm{w}}+\Phi_\text{p}^*\alpha_{\mathrm{p}}),
\end{align}
with
\begin{align}
& \Phi_\text{s}^*(\text{Pe}_*)
= \frac{2}{\pi }  \int_0^{\text{Pe}_*} \frac{du}{\text{Pe}_* \sqrt{1-\left(\frac{u}{\text{Pe}_*}\right)^2}} \Phi_\text{s}(u) \label{PHIStar}
\end{align}
For small arguments, $\Phi_\mathrm{s}^*(u)\simeq \Phi_\mathrm{s}(u)/2$ where the factor one half comes from the quadratic dependence  with the velocity, with $\langle \cos^2(t)\rangle=1/2$. At large Péclet numbers, the force does not depend on the velocity, and the functions $\Phi$ and $\Phi^*$ become equal. The functions $\Phi_\text{p}^*$ and $\Phi_\text{w}^*$ are shown in Fig.~\ref{fig:PHIpStar}.

  \begin{figure}[ht!]
\includegraphics[width=8cm]{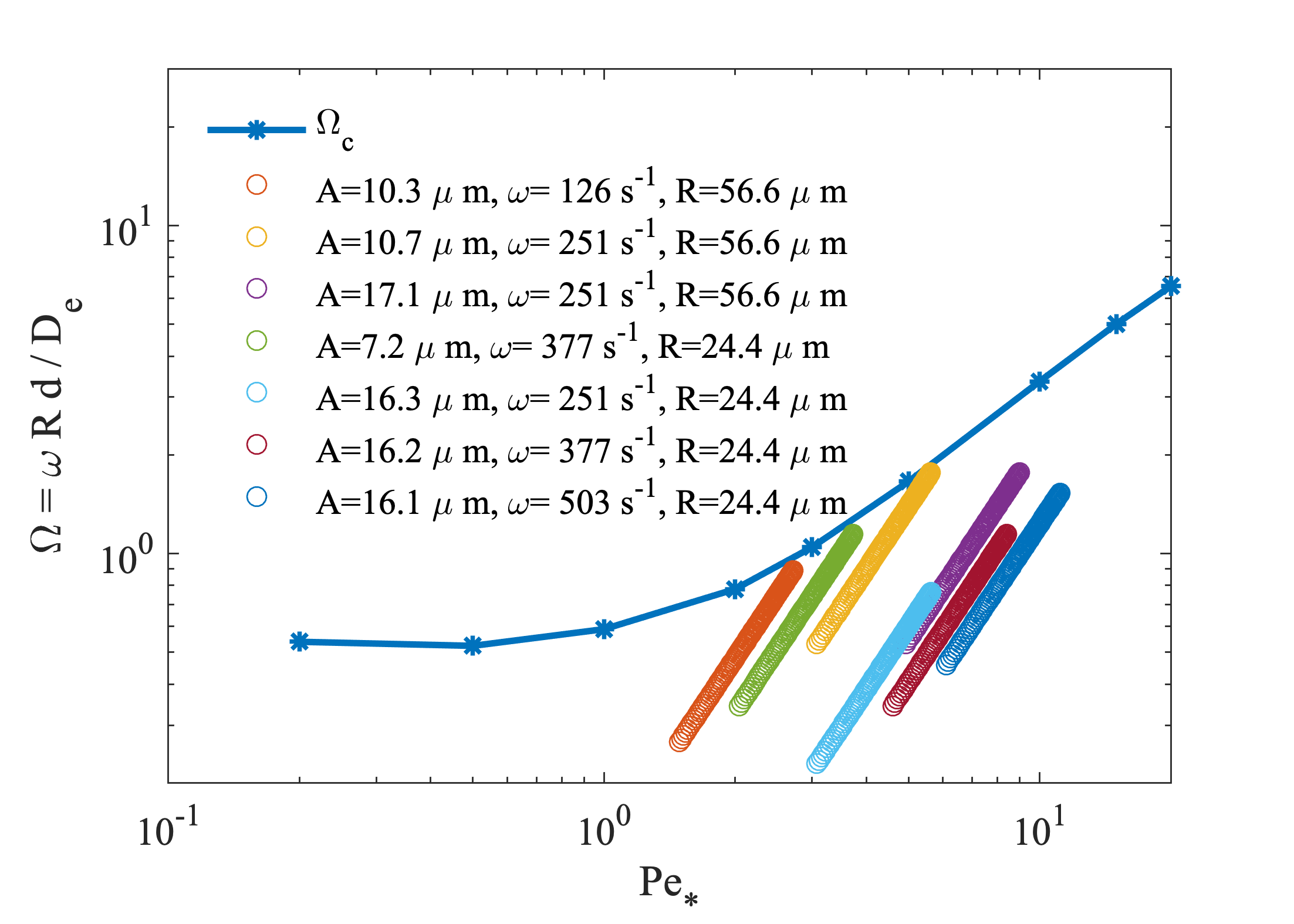}
\caption{$\Omega_c$, defined as the value of $\Omega$ such that the average force differs by less than $15\%$ of the force calculated with the quasi-static hypothesis if $\Omega<\Omega_c$. We report also the value of $\Omega$ for the experimental parameters corresponding to the curves plotted in Fig. 4 of the main text, for distances $d\in[60;200]$ nm. }
\label{FigSafeZone}
\end{figure} 

\section{ Control of the validity of approximations }
\label{SectionControls}

\subsection{Dynamics:  control of the quasi-static approximation}
\label{SectionDyn}

Here we briefly discuss dynamic effects. To evaluate the lift force, we assumed that the lift force can be calculated at each time by using the result of the theory which assumes a stationary state under the velocity (quasi-static assumption). Here we control the validity of this assumption. The slowest varying field is the concentration because its dynamics is governed by diffusion, in comparison the electric field and the hydrodynamic flow adapt much faster to external perturbations. Hence, the only modification in the theory to take into account time dependent effects is to add a term $\partial_tc$ in the equation for $c$. It is useful to use dimensionless variables, we set $\tau=t D_e/l^2$, and $\Omega= \omega l^2/D_e$.

 Equation (\ref{64351}) becomes
\begin{align}
\varepsilon[\partial_\tau c+ \text{Pe} (U_X\partial_X c +U_Y \partial_Yc+U_Z \partial_Z c)]=\partial_Z^2c +\varepsilon \boldsymbol{\nabla}_{\parallel}^2 c ,
\end{align}
where $\text{Pe}(\tau)=\text{Pe}_* \cos(\Omega \tau)$. We still look for solutions under the form $c=c_\infty+c_\infty \lambda^2 /d^{2}      (N+\varepsilon N_1+...),$ where at leading order $N$ does not depend on $Z$.   At next order we have
\begin{align}
&\partial_\tau N+ \text{Pe} \ \ve[U]_\parallel  \cdot\boldsymbol{\nabla}_\parallel  N    = \partial_Z^2N_1 + \boldsymbol{\nabla}_{\parallel}^2 N . 
\end{align}
Integrating over $Z\in[0,H]$ leads to
\begin{align}
H\ \partial_\tau N+&\text{Pe}(\tau) \ \overline{\ve[U]}_\parallel \cdot\boldsymbol{\nabla}_{\parallel} N  =
   \boldsymbol{\nabla}_{\parallel}\cdot (H\boldsymbol{\nabla}_{\parallel}  N ) \nonumber\\
&+2\text{Pe}(\tau)  (\alpha_{\text{w}}S_\perp ^{\text{w}}+\alpha_{\text{p}} S_\perp ^{\text{p}}) \label{EqDyn}.
\end{align}
The instantaneous dimensionless force can then be calculated by using Eq. (\ref{1034}). We have numerically solved Eq. (\ref{EqDyn}) and identified, for each value of the Péclet, the value of $\Omega$ for which the temporally averaged force differs by $15\%$ from the result of our theory using the quasi-static approximation. This value is reported in Fig. \ref{FigSafeZone} for experimental parameters, together with the values of $\Omega$ in all the experiments. From the fact that $\Omega<\Omega_c$ for all experiments, we conclude that the oscillation frequency is low enough to ensure the validity of the quasi-static assumption.

\subsection{Inertial effects}
\label{SectionInertia}
In the Navier-Stokes equation, the inertial term along the $x$ direction can be evaluated as $\rho V_0^2/l$ \cite{cherukat1994inertial,davies2018elastohydrodynamic}, whereas the viscous term in the same direction is of the order of $\eta V_0/d^2$. Hence, 
\begin{align}
\frac{\text{inertial} \ \text{terms}}{\text{viscous} \ \text{terms}}\sim \frac{\rho V_0 d^{3/2}}{\eta \sqrt{R}}< 2\times 10^{-4},
\end{align}
where, for the numerical application, we took $\rho=10^3$ kg.m$^{-3}$, $\eta\sim 10^{-3}$ Pa.s, the smallest $R$ in our experiment $R=56.6\mu$m, the largest velocity in our experiments ($V_0\sim A\omega\sim 8 $mm/s), and $d=200$ nm. Hence, inertial effects are very small in the region near the minimal distance between the sphere and the wall. Furthermore, as mentioned in the main text, the inertial lift for a sphere oscillating in the vicinity of a wall can be evaluated as $1.7\rho V_0^2/2$ \cite{cherukat1994inertial,davies2018elastohydrodynamic}, which is smaller than the smallest observed values of the lift forces for all experimental parameters. We conclude that inertial effects can be neglected in our experiment.  

\subsection{Validity of Cox theory}
Cox's equations are valid when $\lambda$ is small compared to all other lengthscales of the problem. This condition is compatible with the lubrication regime $d\ll R$, as long as one remains in the regime $\lambda\ll d$. Next, Cox theory assumes that the flow $\ve[u]^*$ generated by the motion of ions is small compared to the flow $\ve[u]$ generated by the motion of the wall. In the original derivation of Cox, this is justified by taking the limit of small $\lambda$ when the Hartmann number is implicitly kept constant, which has been argued to be not a physically acceptable limit ~\cite{yariv2011streaming}. However, this assumption can be checked self-consistently in the present case. 
First, we check that the velocity in the bulk is dominated by the velocity imposed by the motion of the wall. Since one has $u\sim V$, whereas $u^*\sim v_s$ is given by Eq. (\ref{CharSlip}), so that
\begin{align}
\frac{u^*}{u}\sim \frac{v_s}{V}\sim \frac{c_\infty k_BT\lambda^2}{\eta D_e} \times\frac{D_e}{l V} \times \left(\frac{\lambda}{d}\right)^2. 
\end{align}
The term $D_e/(l V)$ is of the order one, $c_\infty \lambda^2$ can be identified as of the order of the inverse of the Bjerrum length, and $  k_\mathrm{B}T / (\eta D_e)$ is of the order of the ionic radius. As a consequence, as noted in Ref.~\cite{yariv2011streaming}
$k_\mathrm{B}T c_\infty \lambda^2/(\eta D_e) \simeq \mathcal{O}(1)$, irrespective of the ionic concentration or the viscosity. Since the Péclet is of order one, we conclude that $\vert \ve[u]^*\vert\ll \vert\ve[u]\vert$ in the bulk, justifying that one neglects $\ve[u]^*$ in the equation (\ref{AdvDiffc}) for the bulk ionic concentration. Next, another validity condition of Cox theory is that the flow perpendicular to the surfaces, and at distances $\lambda$ from them, is dominated by the flow induced by the motion of the wall, so that the source terms (\ref{Def_s}) are correctly evaluated. This can be checked by noting that the perpendicular flow at distance $\lambda$ from the surfaces reads $u_\perp\sim V\sqrt{\varepsilon}(\lambda/d)^2$. In the EDL, the flow parallel to the surface due to the motion of ions scales as $v_s$, the perpendicular component can be obtained from the incompressibility condition as $u_\perp^*\sim v_s\lambda/l$. This leads to
\begin{align}
 \frac{u^*_\perp}{u_\perp}\sim  \frac{v_s\lambda/l}{V\sqrt{\varepsilon}(\lambda/d)^2}
 \sim \frac{ k_\mathrm{B}T c_\infty \lambda^2}{\eta D_e}\times \frac{D_e}{l V}\times\frac{\lambda}{d}.
\end{align}
With the same arguments as above, we find that $ u^*_\perp/u_\perp\sim (\lambda/d)\ll 1$, so that we expect that Cox theory can be applied in the present geometry. This remark is consistent with the fact that Cox equations were actually recovered in the case $D_+=D_-$ when the Péclet number is moderate in Ref. \cite{schnitzer2012streaming}.

\bibliographystyle{naturemag}

\end{document}